\documentclass[journal,10pt]{IEEEtran}
\usepackage{amssymb}
\usepackage{amsmath}
\usepackage{cite}
\usepackage{url}
\usepackage{xcolor}
\usepackage{cite,graphicx,amsmath,amssymb}
\usepackage{subfigure}
\usepackage{citesort}
\usepackage{fancyhdr}
\usepackage{mdwmath}
\usepackage{mdwtab}
\usepackage{caption}
\usepackage{amsthm}
\usepackage{setspace}
\usepackage{algorithm}
\usepackage{algorithmic}
\usepackage{makecell}
\usepackage{diagbox}
\usepackage{stfloats}
\usepackage{balance} 
\usepackage{graphicx}
\usepackage{epstopdf}
\usepackage{epsfig}

\newtheorem{remark}{Remark}\newtheorem{theorem}{Theorem}

\newtheorem{lemma}{Lemma}

\newtheorem{corollary}{Corollary}

\allowdisplaybreaks
\makeatletter
\def\ScaleIfNeeded{
	\ifdim\Gin@nat@width>\linewidth \linewidth \else \Gin@nat@width
	\fi } \makeatother
 
\begin{document}
\title{STAR-RIS Assisted SWIPT Systems: Active or Passive?\label{key}}
\author{
	Guangyu~Zhu,
	Xidong~Mu,
	Li~Guo,
	Ao~Huang,
	Shibiao~Xu
	\thanks{Part of this work has been submitted to the IEEE Global Communications Conference, Cape Town, South Africa, December 8-12, 2024.}
	\thanks{Guangyu Zhu, Li Guo, Ao Huang and Shibiao Xu are with the Key Laboratory of Universal Wireless Communications, Ministry of Education, Beijing University of Posts and Telecommunications, Beijing 100876, China, also with the School of Artificial Intelligence, Beijing University of Posts and Telecommunications, Beijing 100876, China, and also with the National Engineering Research Center for Mobile Internet Security Technology, Beijing University of Posts and Telecommunications, Beijing 100876, China (email:\{Zhugy, guoli, huangao, shibiaoxu\}@bupt.edu.cn).}
	\thanks{Xidong Mu is with the School of Electronic Engineering and Computer
		Science, Queen Mary University of London, London E1 4NS, U.K. (e-mail:
		\{xidong.mu\}@qmul.ac.uk).}
}
\maketitle
\begin{abstract}
	A simultaneously transmitting and reflecting reconfigurable intelligent surface (STAR-RIS) assisted simultaneous wireless information and power transfer (SWIPT) system is investigated. Both active and passive STAR-RISs are considered. Passive STAR-RISs can be cost-efficiently fabricated to large aperture sizes with significant near-field regions, but the design flexibility is limited by the coupled phase-shifts. Active STAR-RISs can further amplify signals and have independent phase-shifts, but their aperture sizes are relatively small due to the high cost. To characterize and compare their performance, a power consumption minimization problem is formulated by jointly designing the beamforming at the access point (AP) and the STAR-RIS, subject to both the power and information quality-of-service requirements. To solve the resulting highly-coupled non-convex problem, the original problem is first decomposed into simpler subproblems and then an alternating optimization framework is proposed. For the passive STAR-RIS, the coupled phase-shift constraint is tackled by employing a vector-driven weight penalty method. While for the active STAR-RIS, the independent phase-shift is optimized with AP beamforming via matrix-driven semidefinite programming, and the amplitude matrix is updated using convex optimization techniques in each iteration. Numerical results show that: 1) given the same aperture sizes, the active STAR-RIS exhibits superior performance over the passive one when the aperture size is small, but the performance gap decreases with the increase in aperture size; and 2) given identical power budgets, the passive STAR-RIS is generally preferred, whereas the active STAR-RIS typically suffers performance loss for balancing between the hardware power and the amplification power.
\end{abstract}
	
\section{Introduction}
Over the past years, simultaneous wireless information and power transfer (SWIPT) has emerged as a prominent and highly promising technique in the Internet-of-Things (IoT) network \cite{Zhang_SWIPT,Krikidis_SWIPT,Ponnimbaduge_SWIPT}. Taking full advantage of the potential of radio-frequency (RF) signals to carry both energy and information, SWIPT integrates wireless power transfer (WPT) into conventional wireless information transfer (WIT), thus reconciling the power supply and communication for the network. However, limited by the poor transmission efficiency of RF signals in complex and variable environments, SWIPT will struggle to fulfill the communication and charging requirements of massive devices in the future sixth generation (6G) and beyond networks. To overcome this issue, various technical solutions, e.g., smart antennas\cite{Ding_antennas}, have been proposed. Among them, however, a two-dimensional device consisting of elements with tunable reflection properties, known as reconfigurable intelligent surface (RIS) \cite{RIS_survey}, has attracted the most attention recently. For an RIS, by dynamically adjusting the phase-shift and amplitude of each element, the propagation environment can be \emph{controlled} to enhance desired signals and attenuate unwanted ones. This leads to improved transmission efficiency in terms of spectrum utilization and energy utilization \cite{Huang_EE}. Inspired by this, RIS assisted SWIPT systems have become a topic of great interest \cite{Wu_SWIPT_application}. Nevertheless, the \emph{half-space coverage} caused by the reflecting-only nature of conventional RISs severely limits their implementation and application in practical SWIPT systems. 
	
As a complement, a new RIS concept, namely simultaneously transmitting and reflecting RISs (STAR-RISs), has been suggested in \cite{Mu_star}. On the basis of conventional RISs reflecting-only, the hardware modifications in STAR-RISs allow incident signals to transmit to the opposite side of RISs, thus achieving \emph{full-space} coverage \cite{Liu_360}. Typically, the STAR-RISs referred to in the current literature are passive by default. Although the nearly passive nature of STAR-RISs imposes more stringent requirements on phase-shift configuration\cite{Liu_passive_model}, it also allows STAR-RISs to be cost-efficiently fabricated to large aperture sizes. This scalability provides significant degrees of freedom (DoFs) and a substantial near-field communication region, which are indispensable for addressing the challenging communication environments of future SWIPT systems. On the other hand, a new STAR-RIS model at the cost of additional power consumption, known as the active STAR-RIS, has recently been proposed in \cite{Xu_active_model}. With their ability to reflect, transmit, and amplify signals, active STAR-RISs eliminate the \emph{double fading effect} while maintaining full-space coverage, thus improving signal transmission efficiency. They also feature independent phase-shifts for flexible beamforming design. These characteristics position active STAR-RISs as a potential paradigm to boost SWIPT efficiencies in future IoT networks.
	
\subsection{Prior Works}
\emph{1) Passive RIS/STAR-RIS Assisted SWIPT Systems:} Thanks to their significant improvement in transmission efficiency, RISs/STAR-RISs have been introduced as powerful supports for the study of SWIPT systems. In \cite{Wu_SWIPT_WS}, the authors investigated a weighted sum-power maximization problem for all energy devices (EDs) in a RIS assisted SWIPT system, where the individual information user (IU) signal-to-interference-plus-noise ratio (SINR) constraint was met via joint beamforming. Interestingly, the authors of \cite{Pan_SWIPT_WS} made an exchange of the optimization goal and user constraint in \cite{Wu_SWIPT_WS} and studied a RIS assisted multiple input multiple output SWIPT system. Based on these, to better portray the conflict between IUs and EDs, a multi-objective optimization problem (MOOP) framework was studied by the authors of \cite{Khalili_SWIPT_MOOP}, where beamforming vectors at the access point (AP) and phase-shifts at the RIS were jointly optimized to obtain the fundamental trade-off between sum-rate and total harvested energy. In other respects, under the quality-of-service (QoS) requirements, the authors of \cite{Wu_SWIPT_QoS} proposed to optimize joint active and passive beamforming for transmit power minimization in a multiple RISs assisted SWIPT system. Driven by similar QoS requirements, the transmit power minimization problem was extended for a large-scale RIS aided SWIPT system in \cite{Xu_non_linear}. Furthermore, the authors of \cite{Zargari_SWIPT_NL} studied the max-min fair energy efficient beamforming design for RIS assisted SWIPT systems with a non-linear energy harvesting (EH) model. However, the literature mentioned above strictly requires that all IUs/EDs and transmitters be located on the same side of the RIS. To circumvent this limitation, STAR-RISs were introduced into SWIPT systems in \cite{Yaswanth_STAR-RiS_SWIPT,Zhang_STAR-RIS_SWIPT,Zhu_SWIPT}. More specifically, the authors of \cite{Yaswanth_STAR-RiS_SWIPT} investigated the transmit power minimization problem for a STAR-RIS assisted SWIPT system via a joint beamforming design while guaranteeing the minimum QoS for SWIPT. The authors of \cite{Zhang_STAR-RIS_SWIPT} developed a gamma approximation method to analyze the performance of a STAR-RIS assisted SWIPT system over the Rayleigh fading channel. In addition, in \cite{Zhu_SWIPT}, we investigated the robust resource allocation design for STAR-RIS assisted SWIPT systems under the assumption of imperfect channel state information (CSI), where the max-min rate-energy region was explored by solving a MOOP.
	
\emph{2) Active RIS Assisted SWIPT Systems:} To mitigate the double fading effect caused by passive RISs, active RISs have become the new favorites of auxiliary SWIPT systems and have been studied in \cite{Gao_active,Ren_active_SWIPT,Yaswanth_active_SWIPT,Zargari_versus}. In particular, the authors of \cite{Gao_active} studied two joint beamforming optimization problems with different practical objectives when introducing an active RIS into a SWIPT system. Following this, the authors of \cite{Ren_active_SWIPT} considered an active RIS aided SWIPT downlink system, and aimed to maximize the downlink weighted sum-rate subject to harvested energy constraints in each ED. Under the statistical channel estimation error constraint, the authors of  \cite{Yaswanth_active_SWIPT} investigated the transmit power minimization problem through joint beamforming design for an active RIS assisted SWIPT system. Further considering a realistic piecewise nonlinear EH model, the authors of \cite{Zargari_versus} studied a robust resource allocation strategy in an active RIS assisted multiuser SWIPT system.
	
\subsection{Motivations and Contributions}
It can be observed that despite extensive research on passive STAR-RIS assisted SWIPT systems, there remains a notable gap in exploring active STAR-RISs in this context. This lack of exploration motivates our initial investigation. As we delved into this topic, an interesting question arises, \emph{Which is better in SWIPT systems, active or passive STAR-RISs?} While the authors of \cite{Zhi_versus,Zhang_versus,You_versus} have addressed similar questions for RIS assisted conventional communication systems, their findings may not directly apply to the more complicated STAR-RIS assisted SWIPT systems due to the following reasons:
	
\textbf{Complex phase-shift adjustment constraints:} Compared to conventional RISs, STAR-RISs face more complex phase-shift adjustment constraints. For passive STAR-RISs, in order to ensure the losslessness of the incident signal, their hardware design imposes an inherent constraint on phase-shift coupling between transmission and reflection. However, active STAR-RISs can circumvent this limitation by incorporating amplifier elements. As a result, this leads to a significant difference in phase-shift adjustment between passive and active STAR-RISs, which is absent in conventional RISs. 
	
\textbf{Signal processing needs for EDs vs. IUs:} Compared to conventional communication systems, SWIPT systems cater to a more diverse set of users. These include not only IUs that need to decode the received signal but also EDs that only need to capture energy from the received signal. In this context, the additional thermal noise introduced by active STAR-RISs, which would typically be eliminated in conventional communication systems, can actually be beneficial for EUs in SWIPT systems. Therefore, this conflict in user requirements adds another layer of complexity to the comparison between passive and active STAR-RISs in SWIPT systems.
	
\textbf{Near- and far-field effects:} In particular, passive STAR-RISs typically require a significantly larger size compared to their active counterparts under the same power budget. Consequently, a notable extension of the Rayleigh distance occurs with passive STAR-RISs, leading some users to transition from the far-field region to the near-field region. This shift broadens beamforming considerations from the angular domain to the distance domain, thereby highlighting the inherent differences in beamforming design between passive and active STAR-RISs, an aspect not extensively investigated in previous works.

To the best of our knowledge, no existing studies have comprehensively addressed the above challenges to answer the question posed. Motivated by this, in this paper, we study the comparison between active and passive STAR-RISs to identify a more suitable technical solution for future SWIPT systems. In particular, a multi-user SWIPT system is considered, where the STAR-RIS is introduced to assist a multi-antenna AP to simultaneously transmit information and power to single-antenna IUs and EDs, respectively. To visualize the comparison, we use the AP power consumption under user QoS constraints as a metric and develop a performance comparison between active and passive STAR-RISs in two different contexts. It suggests that \emph{in SWIPT systems, active STAR-RISs are superior for the same aperture size, while passive STAR-RISs are more robust for the same power budget}. The main contributions of this paper are summarized as follows:
\begin{itemize}
	\item We investigate the multi-user SWIPT systems assisted by both passive STAR-RIS and active STAR-RIS scenarios. To characterize their performance, we formulate a joint beamforming design problem for both models to minimize the AP power consumption, subject to user QoS requirements. On this basis, we conduct their performance comparison under the same conditions of either aperture size or power budget. 
	\item For the passive STAR-RIS, we first introduce an equivalent but more tractable expression to replace the coupled phase-shift constraint. Then, we employ the alternating optimization (AO) framework to decompose the original problem into two subproblems, i.e., AP beamforming design and STAR-RIS beamforming design, and effectively solve them in an iterative manner. We utilize the semidefinite relaxation (SDR) method to optimize the AP beamforming. In addition, we propose a weight penalty method to find high-quality solutions for STAR-RIS beamforming with coupled phase-shift.
	\item For the active STAR-RIS, we first consider phase-shift optimization and amplitude optimization in the beamforming design separately. Then, we develop an efficient algorithm by exploiting the SDR method and successive convex approximation (SCA) to jointly optimize the AP beamforming and STAR-RIS phase-shift, where the rank-one constraint is guaranteed by Gaussian randomization. Based on this, we use convex optimization techniques to update the STAR-RIS amplitude in each iteration.
	\item Our numerical results demonstrate that: 1) the introduction of STAR-RISs leads to significant power consumption savings in SWIPT systems; 2) at the same dimensions, active STAR-RISs exhibit superior performance over passive ones, especially when the dimension is small;
 	and 3) with the same power budget constraint, passive STAR-RISs achieve more consistent performance than active STAR-RISs, as they avoid the trade-off between hardware power and amplification power.
\end{itemize}

\subsection{Organization and Notation}
The rest of the paper is structured as follows: Section II presents an introduction to the system model and formulates problems for both passive and active STAR-RISs. Next, efficient algorithms are developed to solve the formulated problems for passive and active STAR-RISs in Section III and Section IV, respectively. Section V shows the numerical results and the corresponding discussions. Finally, the paper concludes with Section VI.
	
\emph{Notations}: Scalars, vectors, and matrices are denoted by lower-case, bold lower-case letters, and bold upper-case letters, respectively. $(\cdot)^T$ and $(\cdot)^H$ denote the transpose and conjugate transpose, respectively. $\|\cdot\|$, $\|\cdot\|_2$, and $\|\cdot\|_F$ denote the norm, spectral norm, and Frobenius norm, respectively. $\mathrm{Tr}(\cdot)$ and $\mathrm{Rank}(\cdot)$ denote the trace and rank of the matrices. $[\cdot]_{m,n}$ denotes the $(m,n)$-th element of the matrix. Besides, $\mathrm{diag}(\cdot)$ denotes the diagonalization operation on vectors. $\mathbb{C}^{M \times N}$ denotes the space of $M \times N$ complex valued matrices. $\mathcal{CN}(\mu, \sigma^2)$ represents the distribution of a circularly symmetrical complex Gaussian random variable with a mean of $\mu$ and a variance of $\sigma^2$. 
The function mod$(a,b)$ calculates the remainder when $a$ is divided by $b$, and $\lfloor a,b\rfloor$ denotes the floor function of the result of dividing $a$ by $b$.  $\otimes$ represents the Kronecker product.
	
\section{System Model and Problem Formulation}
\begin{figure}
	\setlength{\abovecaptionskip}{0cm}   
	\setlength{\belowcaptionskip}{0cm}   
	\setlength{\textfloatsep}{7pt}
	\centering
	\includegraphics[width=3.0in]{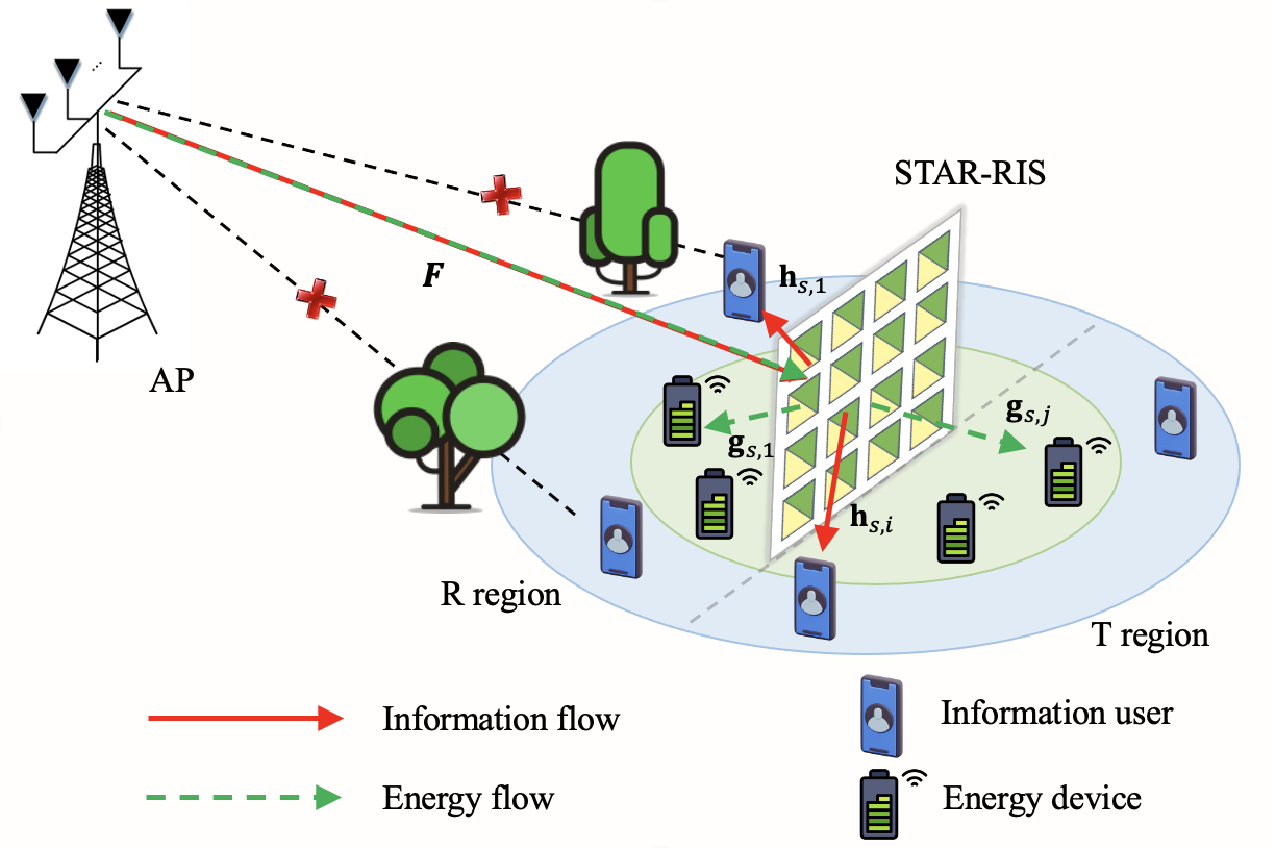}
	\caption{Illustration of the SATR-RIS assisted SWIPT system}
	\label{system}
\end{figure}
A STAR-RIS assiste SWIPT system is illustrated in Fig. \ref{system}, where two groups of IUs and EDs, denoted by $\mathcal{K_I}=\{1,\cdots,K_I\}$ and $\mathcal{K_E}=\{1,\cdots,K_E\}$, are assisted by a STAR-RIS in exchanging information and recharging with the AP, respectively. In particular, the STAR-RIS contains $M$ tunable elements, the AP is equipped with $N$ antennas, while all IUs and EDs are in a single-antenna configuration. With the introduction of the STAR-RIS, the full communication region is partitioned into two parts: the transmission (T) region and the reflection (R) region, where all IUs and EDs are randomly located. Affected by obstacles in the real communication environment, such as tall buildings, trees, etc., we assume that the direct links between the AP and all users are blocked \cite{Mu_star,Zhu_SWIPT}. As a result, communication for users in blind spots is supported by the STAR-RIS enabled transmission and reflection links. In this case, the channel coefficients from the AP to the STAR-RIS, from the STAR-RIS to IU $i$, and from the STAR-RIS to ED $j$ are denoted by $\mathbf{F}\in\mathbb{C}^{M\times N}$, $\mathbf{h}_{s,i}\in \mathbb{C}^{M\times1}$, and $\mathbf{g}_{s,j}\in\mathbb{C}^{M\times1}$, respectively. Besides, note that the channel estimation methods proposed in \cite{Xu_estimation} and \cite{Wu_estimation} are applicable for efficient estimation of the WPT and WIT channels, respectively. Therefore, we assume that the CSI of all links at the AP is perfectly known in order to explore the fundamental design insights.
	
\subsection{Channel Models}
Given the coverage characteristics of the STAR-RIS, its advantages are better exploited by deploying it between users rather than in close proximity to the AP. In this context, the channels between the AP and the STAR-RIS, and between the STAR-RIS and users, can generally be modeled as follows:
	
\emph{1) AP-STAR-RIS Channel Model:} For the uniform linear array (ULA)-type AP and uniform planar array (UPA)-type STAR-RIS, we adopt the commonly used geometric model\cite{Geometric_model} to fit the channel between them as 
\begin{align}\label{channel_F}
	\mathbf{F}=\sqrt{\frac{ MN}{L}}\sum_{l=1}^{L}\alpha_l\mathbf{a}_{\textup{S}}\left(\varphi^A_l,\psi^A_l\right)\mathbf{a}^H_\textup{A}\left(\varphi^D_l\right),
\end{align}
where $L$ denotes the number of paths between the AP and the STAR-RIS and $\alpha_l$ represents the path loss of the $l$-th path. In particular, we consider that $l=1$ is the dominant line-of-sight (LoS) link, and $l=2,\cdots,L$ denote the non-line-of-sight (NLoS) links. Besides, $\varphi_l^A$ and $\psi^A_l$ denote the azimuth and elevation angle of arrival (AoA) associated with the STAR-RIS, respectively, and $\varphi_l^D$denotes the angle of departure (AoD) of the AP. As a result, the corresponding array response vectors are denoted by $\mathbf{a}_\textup{S}\left(\varphi^A_l,\psi^A_l\right)$ for the STAR-RIS and $\mathbf{a}_\textup{A}\left(\varphi^D_l\right)$ for the AP. Supposing the STAR-RIS is a UPA of size $M=M_x\times M_z$, the array response vector $\mathbf{a}_\textup{S}\left(\varphi^A_l,\psi^A_l\right)$ is given by
\begin{align}\label{channel_AoA}
	\mathbf{a}_\textup{S}\left(\varphi^A_l,\psi^A_l\right)&=\frac{1}{\sqrt{M}}[1,\cdots,e^{-j\frac{2\pi d}{\lambda_c}\left(M_x-1\right)\sin\varphi_l^A\sin\psi_l^A}]^T \nonumber \\
	&\otimes [1,\cdots,e^{-j\frac{2\pi d}{\lambda_c}\left(M_z-1\right)\cos\psi_l^A}]^T,
\end{align}
where $\lambda_c$ and $d$ denote the wavelength and the element spacing, respectively. Similarly, for the ULA-type AP with $N$ antennas, the array response vector $\mathbf{a}_\textup{A}\left(\varphi^D_l\right)$ is given by
\begin{align}\label{channel_AoD}
	\mathbf{a}_\textup{A}\left(\varphi^D_l\right)=\frac{1}{\sqrt{N}}[1,\cdots,e^{-j\frac{2\pi d}{\lambda_c}\left(N-1\right)\sin\varphi_l^D}]^T.
\end{align}

\begin{figure}
	\setlength{\abovecaptionskip}{0cm}   
	\setlength{\belowcaptionskip}{0cm}   
	\setlength{\textfloatsep}{7pt}
	\centering
	\includegraphics[width=3.0in]{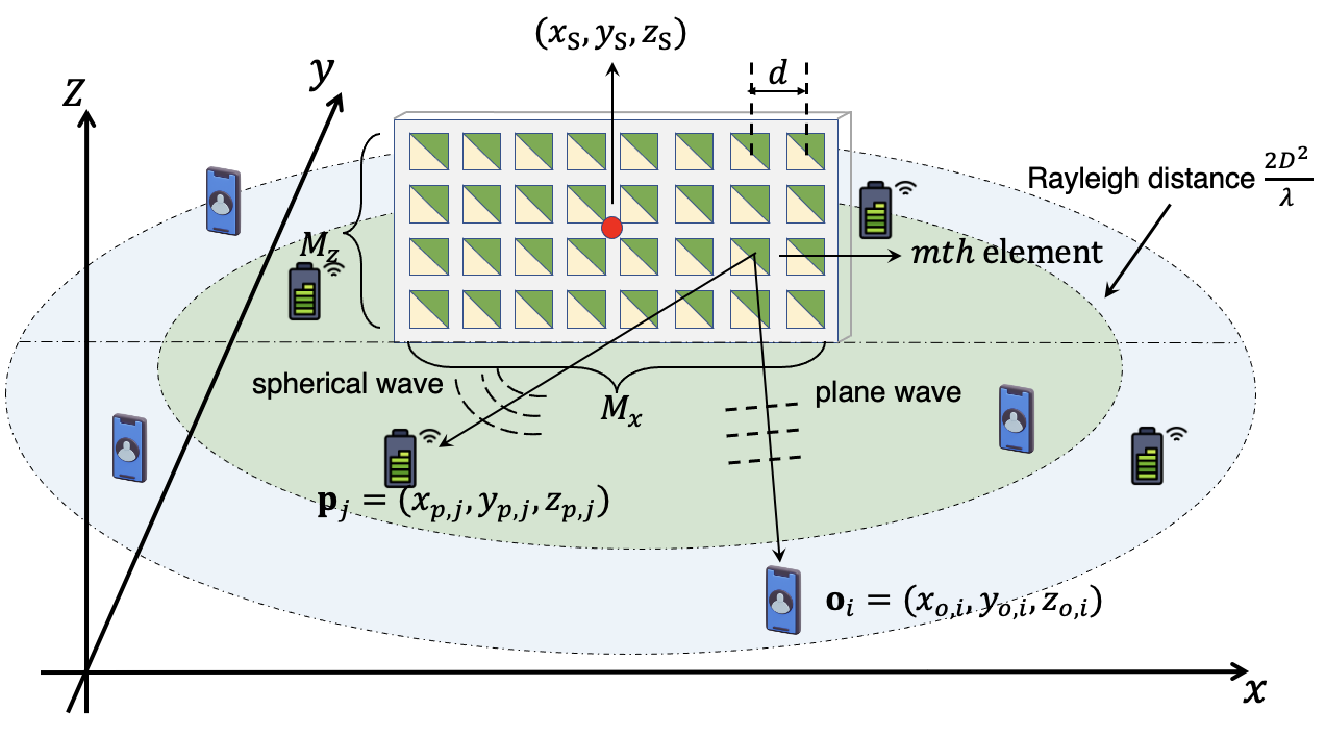}
	\caption{Illustration of the near- and far-field channel models}
	\label{model}
\end{figure}
\emph{2) STAR-RIS-Users Channel Model:} Considering the case of STAR-RIS with the large size involved in this study, the conventional far-field channel model may not accurately cover all user channels. To address this issue, we refine a more complex but flexible channel model to better match the links from the STAR-RIS to users in different locations. As depicted in Fig. \ref{model}, we consider a three-dimensional (3D) topology, where the coordinates at the center of STAR-RIS are denoted as $\mathbf{u}_s=[x_\textup{S},y_\textup{S},z_\textup{S}]^T$. On this basis, the coordinates of the $m$-th element of the STAR-RIS are expressed as
\begin{align}
	\mathbf{u}_m=[x_\textup{S}+\left(i_x(m)-x_m\right)d,y_\textup{S},z_\textup{S}+\left(i_z(m)-z_m\right)d]^T\!\!, 
\end{align}
where $i_x(m)\!=\!\textup{mod}\left(m,M_x\right)$ if $\textup{mod}\left(m,M_x\right)\! \ne \!0$ otherwise $M_x$ and $i_z(m)=\lfloor m,M_x \rfloor \!+\!1$ if $\textup{mod}\left(m,M_x\right) \ne 0$ otherwise $\lfloor m,M_x \rfloor$ denote the indexes of the $m$-th element in the $x$-axis and $z$-axis on the STAR-RIS, respectively. Besides, $x_m=\frac{M_x+1}{2}$ and $z_m\!\!=\!\!\frac{M_z+1}{2}$. 
Accordingly, its Rayleigh distance can be derived from $\frac{2D^2}{\lambda_c}$, where $D$ is the STAR-RIS aperture size. At this point, when the user is beyond this threshold, it will receive an approximate plane wave from the STAR-RIS, allowing the channel to be simulated using the far-field model \cite{You_Near_Far}. Conversely, when the user is within this threshold, it receives a spherical wave, making the near-field model a more accurate representation of the channel \cite{Liu_near_field}.
Using the energy user as an example, suppose that the coordinates of ED $j$ are $\mathbf{p}_j=[x_{p,j},y_{p.j},z_{p,j}]^T$ and fall within the Rayleigh distance. Then, the near-field channel model between the STAR-RIS and ED $j$ is given by \cite{Cui_near_field}
\begin{align}
	\mathbf{g}_{s,j}=[\alpha_{j,1}e^{-j\frac{2\pi r_{j,1}}{\lambda_c}},\cdots,\alpha_{j,M}e^{-j\frac{2\pi r_{j,M}}{\lambda_c}}]^T,
\end{align}
where $\alpha_{j,m}=\frac{\lambda_c}{4\pi r_{j,m}}$ and $r_{j,m}=\|\mathbf{u}_m-\mathbf{p}_j\|_2$ represent the free space path loss coefficient and the distance between the $m$-th element of the STAR-RIS and ED $j$, respectively. However, when it lies outside the Rayleigh distance, its channel will transform into a far-field model, shown as
\begin{align}
	\mathbf{g}_{s,j}=\alpha_j\sqrt{M}\mathbf{a}_\textup{S}\left(\varphi^D_j,\psi^D_j\right),
\end{align}
where $\alpha_j=\frac{\lambda_c}{4\pi r_j}$ and $r_j=\|\mathbf{u}_s-\mathbf{p}_j\|_2$. Besides, $\mathbf{a}_\textup{S}\left(\varphi^D_j,\psi^D_j\right)$ has a similar form to \eqref{channel_AoA}, where $\varphi^D_j$ and $\psi^D_j$ denote the azimuth and elevation AoD associated with the STAR-RIS, respectively. Note that we ignore the NLoS paths from the STAR-RIS to EDs and IUs here due to their weak power \cite{You_Near_Far}.
	
Along the same line, assume further that the coordinates of IU $i$ are denoted as $\mathbf{o}_i=[x_{o,i},y_{o,i},z_{o,i}]^T$. Then, its channel model can be represented as 
\begin{align}
	\mathbf{h}_{s,i}\!\!=\!
	\begin{cases}
		[\alpha_{i,1}e^{-j\frac{2\pi r_{r,1}}{\lambda_c}},\!\cdots,\!\alpha_{i,M}e^{-j\frac{2\pi r_{i,M}}{\lambda_c}}]^T,\!\! \!&\!\textup{near-field},\\
		\alpha_i\sqrt{M}\mathbf{a}_\textup{S}\left(\varphi^D_i,\psi^D_i\right), &\textup{far-field},
	\end{cases}
\end{align}
where $\alpha_{i,m}=\frac{\lambda_c}{4\pi r_{i,m}}$, $r_{i,m}=\|\mathbf{u}_m-\mathbf{o}_i\|_2$, $\alpha_i=\frac{\lambda_c}{4\pi r_i}$, and $r_i=\|\mathbf{u}_s-\mathbf{o}_i\|_2$.
\begin{remark}
	\textup{Indeed, the near-field model considers both the angular and distance domains in the beamforming design, whereas the far-field model starts with the angular domain only. Therefore, the latter may be regarded as a special case of the former. For the convenience of subsequent analysis, we will directly denote the channels between the STAR-RIS and EDs or IUs as $\mathbf{g}_{s,j}$ or $\mathbf{h}_{s,i}$, without specifically distinguishing between near-field and far-field models.}
\end{remark}
\subsection{STAR-RIS Signal Models}
In this paper, we consider both the passive and active models for the STAR-RIS employing energy splitting protocol.
	
\emph{1) Passive STAR-RIS Model:} As mentioned in \cite{Liu_passive_model}, the incident signal can be manipulated to undergo both transmission and reflection through circuit adjustments for each STAR-RIS element. Let $s_m$ denote the incident signal of the $m$-th element, then the transmitted and the reflected signal in this element can be denoted as $t_m=\sqrt{\beta^t_m}e^{j\theta^t_m}s_m$ and $r_m=\sqrt{\beta^r_m}e^{j\theta^r_m}s_m$, respectively, where $\beta^s_m\in[0,1]$ and $\theta^s_m\!\in\! [0,2\pi)$, $s\!\in\!\{t,r\}$ represent the amplitude and phase-shift adjustments imposed on the incident signal. Based on this, the transmission- and reflection-coefficient matrices of the passive STAR-RIS can be given by $\mathbf{\Theta}^{\textup{pass}}_t=\textup{diag}\left(\sqrt{\beta_1^{t}}e^{j\theta^t_1},\cdots,\sqrt{\beta_M^{t}}e^{j\theta^t_M}\right)$ and $\mathbf{\Theta}^{\textup{pass}}_r=\textup{diag}\left(\sqrt{\beta_1^{r}}e^{j\theta^r_1},\cdots,\sqrt{\beta_M^{r}}e^{j\theta^r_M}\right)$, respectively. It should be noted that given the law of energy conservation and the realistic electric and magnetic impedances inherent to the STAR-RIS, the following constraint implies that $\beta^t_m+\beta^r_m=1$ and $\cos\left(\theta^t_m-\theta^r_m\right)=0, \forall m \in\mathcal{M} \triangleq \{1,\cdots,M\}$ consistently upheld within each individual element. In addition, due to the nearly passive nature of the passive STAR-RIS, its power consumption can be modeled as $P_\textup{pass}=MP_c$ \cite{Zhi_versus}, where $P_c$ denotes the circuit power consumption of each element used to control the phase-shift adjustment.
	
\emph{2) Active STAR-RIS Model:} As shown in \cite{Xu_active_model}, the active STAR-RIS construction is based on the hardware of the passive STAR-RIS, with the addition of a reflection-type amplifier for each element. Therefore, the active STAR-RIS functions not only in the transmission and reflection of incident signals, but also in their effective amplification. Let $\mathbf{A}=\textup{diag}\left(a_1,\cdots,a_M\right)$ denote the amplification matrix for the active STAR-RIS, where $a_m\geq0$ is the amplification factor of the $m$-th element. Accordingly, the complete transmission- and reflection-coefficient matrices for the active STAR-RIS can be given by $\mathbf{\Theta}^{\textup{act}}_t=\mathbf{A}\mathbf{\Theta}_t$ and $\mathbf{\Theta}^{\textup{act}}_r=\mathbf{A}\mathbf{\Theta}_r$, respectively, where $\mathbf{\Theta}_s=\textup{diag}\left(\sqrt{\beta_1^{s}}e^{j\theta^s_1},\cdots,\sqrt{\beta_M^{s}}e^{j\theta^s_M}\right)$, with $\beta^s_m \in [0,1], \theta^s_m \in[0,2\pi), s\in\{t,r\}$. In particular, with the introduction of amplifiers, the flexible hardware design can effectively circumvent the coupling of phase-shift. As such, the stringent restrictions on T\&R within each active STAR-RIS element are solely confined to amplitude and considered by $\beta^t_m+\beta^r_m=1, \forall m \in \mathcal{M}$. Besides, since the active STAR-RIS amplifies the signal, its power consumption model is extended on the basis of the passive STAR-RIS as $P_\textup{act}=M(P_c+P_b)+P_a$, where $P_b$ is the biasing power used by the amplifier in each element and $P_a$ denotes the allowable power budge for amplifying signals of the STAR-RIS.
\begin{remark}
	\textup{Since the power consumption of the passive STAR-RIS comes only from the circuit, it can have a much larger size than the active STAR-RIS as the STAR-RIS power budget increases. In this case, the near-field effect is more pronounced for the passive STAR-RIS than its active counterpart, but the inherent phase-shift coupling is also more complicated.}
\end{remark}
	
\subsection{Signal Transmission Models}
In this paper, we adopt linear transmit precoding at the AP, which assigns a dedicated beam, denoted by $\mathbf{w}_i\in\mathbb{C}^{N\times1}$ and $\mathbf{v}_j\in\mathbb{C}^{N\times1}$, for each IU $i$ and each ED $j$, respectively. Accordingly, the transmitted signal at the AP is expressed as 
\begin{align}
	\mathbf{x}=\sum_{i\in \mathcal{K_I}}\mathbf{w}_ix^I_i+\sum_{j\in\mathcal{K_E}}\mathbf{v}_jx^E_j,
\end{align}
where $x_i^I \sim \mathcal{CN}(0,1)$ is the information-bearing signal for IU $i$, and $x_j^E$ with $\mathbb{E}(|x_j^E|^2)=1$ is the energy-carrying signal for ED $j$. For the sake of simplicity, we assume that the energy beam and information beam are independent. Therefore, the transmit power of the AP can be expressed as 
\begin{align}	P_T=\mathbb{E}\{\mathbf{x}^H\mathbf{x}\}=\sum_{i\in\mathcal{K_I}}\|\mathbf{w}_i\|^2+\sum_{j\in\mathcal{K_E}}\| \mathbf{v}_j \|^2.
\end{align}
	
\emph{1) Passive STAR-RIS Assisted SWIPT:} When enabling a passive STAR-RIS to assist in communication, the received signal at IU $i$ can be expressed as 
\begin{align}\label{received_IU}
	y^I_{\textup{pass},i}=\mathbf{h}^H_{s,i}\mathbf{\Theta}^{\textup{pass}}_{s_i}\mathbf{F}\mathbf{x}+n_i, \forall i \in \mathcal{K_I},
\end{align}
where $s_i\in\{t,r\}$ indicates the T/R region where IU $i$ is located. $n_i \in \mathcal{CN} \left(0,\sigma^2\right)$ denotes the additive white Gaussian noise at IU $i$. Given that the energy beam contains no information, we assume it can be successfully decoded and eliminated by IUs. As a result, the achievable SINR can be expressed as
\begin{align}
	\textup{SINR}_{\textup{pass},i}=\frac{|\mathbf{q}^H_{s_i}\mathbf{H}_i\mathbf{w}_i|^2}{\sum_{k\in\mathcal{K_I},k\ne i}|\mathbf{q}^H_{s_i}\mathbf{H}_i\mathbf{w}_k|^2+\sigma^2}, 
\end{align}
where $\mathbf{H}_i=\mathrm{diag}\left(\mathbf{h}_{s,i}\right)\mathbf{F} \in \mathbb{C}^{M \times N}$ denotes the cascaded channel from the AP to IU $i$ via the STAR-RIS. $\mathbf{q}_{s_i}=[\sqrt{\beta^{s_i}_1}e^{j\theta^{s_i}_1},\cdots,\sqrt{\beta^{s_i}_M}e^{j\theta^{s_i}_M}]^T \in \mathbb{C}^{M\times 1}, s_i\in\{t,r\}$ represents the passive STAR-RIS tuning vector.

On the other hand, the received signal by ED $j$ can also be expressed in a form similar to \eqref{received_IU} as follows:
\begin{align}\label{received_ED}
	y^E_{\textup{pass},j}=\mathbf{g}^H_{s,j}\mathbf{\Theta}^{\textup{pass}}_{s_j}\mathbf{F}\mathbf{x},+n_j, \forall j \in \mathcal{K_E},
\end{align}
where $s_j\in\{t,r\}$ and $n_j\in \mathcal{CN}\left(0,\sigma^2\right)$. Different from information users, energy devices do not require detailed decoding of received signals, which results in all signals being desirable to energy devices. In addition, the energy harvesting levels considered in this system can be competely covered by the linear conversion region generated by multi-parallel EH circuits \cite{Ma_linear}. Therefore, for the simplicity of the study, we adopt the linear EH model \cite{Chen_MEC} to express the received RF energy/power\footnote{In this paper, we use the unit time of 1 second to measure system performance. Thus, the terms “power” and “energy” are interchangeable.} at ED $j$ as follows:
\begin{align}\label{P_pass}
	P_{\textup{pass},j}=\eta\left(\sum_{i\in\mathcal{K_I}}\big|\mathbf{q}_{s_j}^H\mathbf{G}_j\mathbf{w}_i\big|^2+\sum_{k\in\mathcal{K_E}}\big|\mathbf{q}_{s_j}^H\mathbf{G}_j\mathbf{v}_k\big|^2\right),
\end{align}
where $\mathbf{G}_j=\mathrm{diag}\left(\mathbf{g}_{s,j}\right)\mathbf{F} \in \mathbb{C}^{M\times N}$ denotes the cascaded channel from the AP to ED $j$, $\eta$ is the energy conversion efficiency.  Note that the power of the noise is much lower than the received signal, which leads to its absence in \eqref{P_pass}.
	
2) \emph{Active STAR-RIS Assisted SWIPT}: Unlike the passive transmission process described above, the implementation of active STAR-RIS also amplifies the signal accordingly. In this way, the thermal noise generated by amplifiers becomes an unavoidable part of the transmission process \cite{Long_active}. Therefore, the signals received by IU $i$ and ED $j$ from the AP with the help of active STAR-RIS are respectively represented as
\begin{align}
	y_{\textup{act},i}^I=\mathbf{h}^H_{s,i}\mathbf{\Theta}^{\textup{act}}_{s_i}\mathbf{F}\mathbf{x}+\mathbf{h}^H_{s,i}\mathbf{\Theta}^{\textup{act}}_{s_i}\mathbf{z}+n_i, \forall i \in \mathcal{K_I}, \\
	y_{\textup{act},j}^E=\mathbf{g}_{s,j}^{H}\mathbf{\Theta}^{\textup{act}}_{s_j}\mathbf{F}\mathbf{x}+\mathbf{g}^H_{s,j}\mathbf{\Theta}^{\textup{act}}_{s_j}\mathbf{z}+n_j, \forall j \in \mathcal{K_E}.
\end{align}
It may be noted that $\mathbf{z}\in\mathcal{CN}\left(0,\sigma^2_z\mathbf{I}_{M\times M}\right)$ is the thermal noise introduced by the active STAR-RIS with the power $\sigma^2_z$. Following this, the achievable SINR for IU $i$ can be expressed as 
\begin{align}
	\textup{SINR}_{\textup{act},i}\!=\!\frac{|\widetilde{\mathbf{q}}_{s_i}^H\mathbf{H}_i\mathbf{w}_i|^2}{\sum_{k\in\mathcal{K_I},k\ne i}\!|\widetilde{\mathbf{q}}_{s_i}^H\mathbf{H}_i\mathbf{w}_k|^2\!+\!\sigma^2_z|\mathbf{h}^H_{s,i}\widetilde{\mathbf{q}}_{s_i}|^2\!\!+\!\sigma^2},
\end{align}
where $\widetilde{\mathbf{q}}_{s_i}=[a_1\sqrt{\beta^{s_i}_1}e^{j\theta^{s_i}_1}, \cdots, a_M\sqrt{\beta^{s_i}_M}e^{j\theta^{s_i}_M}]^T=\mathbf{A}\mathbf{q}_{s_i} \in \mathbb{C}^{M\times 1}, s_i\in\{t,r\}$ represents the active STAR-RIS tuning vector. Along a similar line, the energy harvesting by ED $j$ can be denoted by $P_{\textup{act},j}$, shown as 
\begin{align}
	P_{\textup{act},j}\!\!=\!\eta\!\left(\!\sum_{i\in\mathcal{K_I}}\!\!\big|\widetilde{\mathbf{q}}_{s_j}^H\mathbf{G}_j\mathbf{w}_i\big|^2\!\!+\!\!\!\sum_{k\in\mathcal{K_E}}\!\!\big|\widetilde{\mathbf{q}}_{s_j}^H\mathbf{G}_j\mathbf{v}_k\big|^2\!\!+\!\!\sigma^2_z|\mathbf{g}^H_{s,j}\widetilde{\mathbf{q}}_{s_j}|^2\!\right).
\end{align}

Due to the amplification of the incident signal, the energy consumption for signal processing of the active SATR-RIS is significantly larger than that of the passive STAR-RIS, and therefore it cannot be ignored. Assume that the maximum amplification power budget $P_a$ is available to the active STAR-RIS. We can have
\begin{align}\label{Power_constraint}
	&\!\!\!\sum_{s \in \{t,r\}}\!\!\!\!\left(\!\sum_{i\in \mathcal{K_I}}\!\!||\mathbf{A}\mathbf{\Theta}_{s}\mathbf{F}\mathbf{w}_i||^2\!\!+\!\!\!\sum_{j\in\mathcal{K_E}}\!\!\!||\mathbf{A}\mathbf{\Theta}_{s}\mathbf{F}\mathbf{v}_j||^2\!\!+\!\sigma^2_z\|\mathbf{A}\mathbf{\Theta}_{s}\|_F^2\! \!\!\right)\nonumber \\
	&\!\!\!=\!\!\!\!\!\sum_{s \in \{t,r\}}\!\!\!\left(\!\sum_{i\in \mathcal{K_I}}\!\!|\widetilde{\mathbf{q}}^H_{s}\mathbf{F}\mathbf{w}_i|^2\!\!+\!\!\!\sum_{j\in\mathcal{K_E}}\!\!|\widetilde{\mathbf{q}}^H_{s}\mathbf{F}\mathbf{v}_j|^2\!\!+\!\!\sigma^2_z\|\widetilde{\mathbf{q}}_s\|^2 \!\!\right)\!\!\leq \!\!P_a.
\end{align}
	
\subsection{Problem Formulation}
Our goal is to minimize the power consumption $P_T$ at the AP, subject to the QoS of each IU and ED. In this case, the problem of jointly optimizing the AP beamforming and the STAR-RIS beamforming for both passive and active STARis is formulated as 
\begin{subequations}\label{P1}
	\begin{align}
		&\label{P1_C0}\  \min_{{\mathbf{w}_i},{\mathbf{v}_j},{\mathbf{q}_s}/{\widetilde{\mathbf{q}}_s}} \sum_{i\in\mathcal{K_I}}\left\|\mathbf{w}_i\right\|^2+\sum_{j\in\mathcal{K_E}}\left\|\mathbf{v}_j\right\|^2\\ 
		&\label{P1_C1}  {\rm s.t.} \ \textup{SINR}_{\textup{X},i} \geq \gamma_{th},\forall i \in \mathcal{K_I}, \textup{X} \in \{\textup{pass},\textup{act}\},\\
		&\label{P1_C2} \ \ \quad 	P_{\textup{X},j} \geq P_{th}, \forall j \in \mathcal{K_E}, \textup{X} \in \{\textup{pass},\textup{act}\},\\
		&\label{P1_C3} \ \ \quad  \beta^t_m,\beta^r_m\in[0,1],\ \beta^t_m+\beta^r_m=1, \forall m\in\mathcal{M},\\
		&\label{P1_C4} \ \ \quad  \theta^t_m, \theta^r_m\in [0,2\pi), \forall m \in\mathcal{M},\\
		&\label{P1_C5} \ \ \quad  |\theta^t_m\!\!-\!\theta^r_m\!|\!=\!\frac{\pi}{2}\ \textup{or} \ \frac{3\pi}{2},\forall m\! \in \! \mathcal{M}, \textup{for passive model},\\
		&\label{P1_C6} \ \ \quad  a_m \in [0,a_{\max}], \forall m \in \mathcal{M}, \textup{for active model},\\
		&\label{P1_C7}\ \ \quad  P_{\textup{X}} \leq P_S, \textup{X} \in \{\textup{pass},\textup{act}\},
	\end{align}
\end{subequations}
where constraints \eqref{P1_C1} and \eqref{P1_C2} represent the QoS requirements for each IU and each ED, respectively.\footnote{For the sake of simplicity, we assume that all IUs have the same minimum achievable SINR requirement $\gamma_{th}$ with $\gamma_{th}=2^{R_{th}}-1$, and all EDs have the same minimum harvested power requirement $P_{th}.$} On the one hand, \eqref{P1_C3}-\eqref{P1_C5} jointly characterized the tuning constraints for each passive STAR-RIS element under the coupled T\&R phase-shift. On the other hand, the beamforming design constraints for each active SATR-RIS element can be generalized by \eqref{P1_C3}, \eqref{P1_C4}, and \eqref{P1_C6}, where $a_{\max}$ in \eqref{P1_C6} denotes the amplification coefficient constraint. In addition, $P_S$ in  \eqref{P1_C7} denotes the maximum power budget at the STAR-RIS.

Note that given the number of STAR-RIS elements $M$, the hardware power consumption is also determined. As a result, the constraint \eqref{P1_C7} in the passive STAR-RIS can be directly ignored in the subsequent solution, whereas in the active STAR-RIS, it can be simplified to the form \eqref{Power_constraint}. Nonetheless, for both types of STAR-RISs, the problem \eqref{P1} is now presented in a form that is difficult to solve directly. This is due to the fact that all optimization variables are tightly coupled in the objective function and constraints \eqref{P1_C1} and \eqref{P1_C2}, which leads to a high degree of non-convexity in the problem. Besides, compared to STAR-RISs employing independent phase-shift, the constraints are more stringent in our considered coupled phase-shift models, which inevitably creates more obstacles in the system design. Moreover, the amplification tuning coefficient constraint \eqref{P1_C6} in the active STAR-RIS also introduces new difficulties to optimization. In summary, no existing algorithm can be found to solve this problem. To address this issue, we will explore efficient algorithms in the next two sections.
	
\section{Proposed Solution for Passive STAR-RIS}
It can be observed that the problem \eqref{P1} for the version of the active STAR-RIS just adds the amplification coefficient consideration while ignoring the coupled phase-shift constraint to that of the passive STAR-RIS. Therefore, we prefer to explore the solution for the passive STAR-RIS in this section, and thereafter make appropriate modifications based on it to solve the problem for the active STAR-RIS in the next section.

Compared to conventional passive STAR-RIS assisted SWIPT systems, the coupled phase-shift constraint becomes the main obstacle to the solution of the considered problem. To this end, we first define $\mathbf{q}_t=[q_{t,1}, \cdots, q_{t,M}]^T$ and $\mathbf{q}_r=[q_{r,1}, \cdots, q_{r,M}]^T$, where $q_{s,m}=\sqrt{\beta_{s,m}}e^{j\theta_{s,m}}, s\in\{t,r\}, \forall m \in \mathcal{M}$. Moreover, due to the coupled phase-shift, i.e., $|\theta_{t,m}-\theta_{r,m}|=\frac{\pi}{2}$ or $\frac{3\pi}{2}$, non-zero $q_{t,m}$ and $q_{r,m}$ are orthogonal in complex space. In this way, we can rewrite constraint \eqref{P1_C5} in the following equivalent form:
\begin{align}\label{E1}
	\| q_{t,m}+q_{r,m}\|-\|q_{t,m} - q_{r,m}\| =0, \forall m \in \mathcal{M}.
\end{align}
Note that the above equation also holds when $q_{t,m}$ or $q_{r,m}$ is 0, which means that the $m$-th element only works in R or T mode. In this case, $\theta_{t,m}$ and $\theta_{r,m}$ can be independently adjusted. Inspired by this transformation, the penalty function proves to be a favorable method for resolving this new challenging constraint \eqref{E1}. Thus, we further consider constraint \eqref{E1} by integrating it into the following objective function:
\begin{subequations}\label{P2}
	\begin{align}
		&\label{P2_C0}\  \min_{{\mathbf{w}_i},{\mathbf{v}_j},{\mathbf{q}_s}} \sum_{i\in\mathcal{K_I}}\left\|\mathbf{w}_i\right\|^2+\sum_{j\in\mathcal{K_E}}\left\|\mathbf{v}_j\right\|^2 + \xi \textup{X}_{penalty}\\ 
		&\label{P2_C1} \ \quad {\rm s.t.} \ \eqref{P1_C1}, \eqref{P1_C2}, \eqref{P1_C3}, \eqref{P1_C4},
	\end{align}
\end{subequations}
where $\textup{X}_{penalty}\!=\!|\| q_{t,m}+q_{r,m}\|-\|q_{t,m} \!-\! q_{r,m}\||$ is a non-negative penalty term and $\xi$ is the corresponding penalty coefficient. As such, the original problem \eqref{P1} reduces to a common highly-coupled non-convex form \eqref{P2}. Next, we will decompose it into two simpler subproblems, i.e., AP beamforming design and STAR-RIS beamforming design, to decouple the optimization variables. Then, the AO framework is employed to solve these two subproblems iteratively.
\subsection{AP Beamforming Design}
To begin with, we define $\mathbf{W}_i=\mathbf{w}_i\mathbf{w}^H_i \in\mathbb{C}^{N \times N}$, $\mathbf{V}_j=\mathbf{v}_j\mathbf{v}^H_j \in\mathbb{C}^{N \times N}$, $\mathbf{Q}_s=\mathbf{q}_s\mathbf{q}^H_s \in \mathbb{C}^{M \times M}$ with rank-one restriction. Then, for given STAR-RIS beamforming $\{\mathbf{Q}_s\}$, problem \eqref{P2} is reformulated for the AP beamforming design as follows:
\begin{subequations}\label{P3}
	\begin{align}
		&\label{P3_C0} \  \min_{{\mathbf{W}_i},{\mathbf{V}_j}} \sum_{i\in\mathcal{K_I}}\mathrm{Tr}\left(\mathbf{W}_i\right)+\sum_{j\in\mathcal{K_E}}\mathrm{Tr}\left(\mathbf{V}_j\right)\\ 
		&\label{P3_C1} \  {\rm s.t.} \mathrm{Tr}\!\left(\mathbf{W}_i\mathbf{H}^H_i\!\mathbf{Q}_{s_i}\mathbf{H}_i\right)\! \!- \!\gamma_{th}\!\! \!\!\!\!\! \sum_{k\in\mathcal{K_I},k\ne i}\!\!\!\!\!\!\mathrm{Tr}\!\left(\mathbf{W}_k\mathbf{H}^H_i\mathbf{Q}_{s_i}\!\mathbf{H}_i\right)\!\geq \!\gamma_{th} \sigma^2\!\!, \nonumber\\
		&\ \ \quad \forall i \in \mathcal{K_I},\\
		&\label{P3_C2} \ \quad \sum_{i\in \mathcal{K_I}}\!\!\mathrm{Tr}\left(\!\mathbf{W}_i\mathbf{G}_j^H\mathbf{Q}_{s_j}\mathbf{G}_j\!\right)\!+\!\!\!\sum_{k\in\mathcal{K_E}}\!\!\mathrm{Tr}\left(\!\mathbf{V}_k\mathbf{G}_j^H\mathbf{Q}_{s_j}\mathbf{G}_j\!\right)\! \geq \! P_{th}, \nonumber \\
		&\ \ \quad \forall j \in \mathcal{K_E},\\
		&\label{P3_C3} \ \ \quad \mathbf{W}_i \succeq 0, \mathrm{Rank}\left(\mathbf{W}_i\right)=1, \forall i \in \mathcal{K_I},\\
		&\label{P3_C4} \ \ \quad \mathbf{V}_j \succeq 0, \mathrm{Rank}\left(\mathbf{V}_j\right)=1, \forall j \in \mathcal{K_E}. 
	\end{align}
	\end{subequations}
At this point, problem \eqref{P3} is a standard semidefinite program (SDP) problem featuring non-convex rank-one constraints. To effectively solve it, we employ the SDR method to circumvent these constraints directly. This enables the relaxed problem to be solved using existing solvers such as CVX\cite{cvx}. Indeed, there is no gap between the solution to the relaxed version and the original version for \eqref{P3}. This is because the optimal solutions $\mathbf{W}^*_i$ and $\mathbf{V}^*_j$ obtained by solving the former always satisfy the rank-one constraints considered in the latter, which can be demonstrated through \cite[\textbf{Theorem 1}]{Zhu_SWIPT}. Therefore, the desired transmit beamforming vector can be recovered via the Cholesky decomposition as $\mathbf{W}^*_i=\mathbf{w}^*_i\mathbf{w}^{*H}_i$ and $\mathbf{V}^*_j=\mathbf{v}^*_j\mathbf{v}^{*H}_j$.

\subsection{STAR-RIS Beamforming Design}
For given $\{\mathbf{w}_i, \mathbf{v}_j\}$, the optimization objective of problem \eqref{P2} will degrade to include only the penalty term. Let  $\mathbf{A}_{i,k}=\mathbf{H}_i\mathbf{w}_k\mathbf{w}_k^H\mathbf{H}^H_i$, $\mathbf{B}_{j,k}=\mathbf{G}_j\mathbf{w}_k\mathbf{w}^H_k\mathbf{G}^H_j$ and  $\mathbf{C}_{j,k}=\mathbf{G}_j\mathbf{v}_k\mathbf{v}^H_k\mathbf{G}^H_j \in \mathbb{C}^{M\times M}$, we can recast $\eqref{P2}$ as 
\begin{subequations}\label{P4}
	\begin{align}
		&\label{P4_C0}\min_{\mathbf{q}_s} \quad \xi \sum_{m\in\mathcal{M}}\big| \| q_{t,m}+q_{r,m}\|-\|q_{t,m} - q_{r,m}\| \big|\\ 
		&\label{P4_C1}  {\rm s.t.} \ \mathbf{q}^H_{s_i} \mathbf{A}_{i,i} \mathbf{q}_{s_i} \!\!-\!\! \gamma_{th} \sum^{K_I}_{k\ne i}\mathbf{q}^H_{s_i} \mathbf{A}_{i,k} \mathbf{q}_{s_i}\! \geq\! \gamma_{th} \sigma^2\!, \forall i \!\in\! \mathcal{K_I},\\
		&\label{P4_C2} \ \ \quad \sum_{k=1}^{K_I} \mathbf{q}^H_{s_j} \mathbf{B}_{j,k} \mathbf{q}_{s_j}\!\!+\!\sum_{k=1}^{K_E} \mathbf{q}^H_{s_j} \mathbf{C}_{j,k} \mathbf{q}_{s_j} \!\!\geq \!P_{th}, \forall j \!\in\!\mathcal{K_E},\\
		&\label{P4_C3} \ \quad \ \eqref{P1_C3}.
	\end{align}
\end{subequations}
Nonetheless, the non-convexity of the penalty term makes the reformulated problem \eqref{P4} still difficult to solve. To overcome this challenge, we first introduce a non-negative auxiliary variable $\{Z_m \geq 0\}$ that satisfies $Z_m\geq \| q_{t,m}+q_{r,m}\|-\|q_{t,m} - q_{r,m}\|$ and $Z_m\geq \| q_{t,m}-q_{r,m}\|-\|q_{t,m} + q_{r,m}\|$. Besides, inspired by the SCA technique, we can further transform the right-hand side of the above two inequalities with the corresponding convex approximate upper bound by the first-order Taylor expansion as \eqref{A1} and \eqref{A2}, shown at the top of the next page,
\begin{figure*}[]
	\setlength{\belowdisplayskip}{-1pt}
	\normalsize
	\begin{align}
		&\label{A1}\|q_{t,m}\!\!+q_{r,m}\|\!-\!\|q_{t,m} \!\!- q_{r,m}\| \!\leq \!\| q_{t,m}\!\!+q_{r,m}\| \!-\! \|q^{(l)}_{t,m}\!\!-q^{(l)}_{r,m}\|\!-\!\textup{Re} \left\{\frac{\left(q^{(l)}_{t,m}\!\!-q^{(l)}_{r,m}\right)^*}{\|q^{(l)}_{t,m}\!\!-q^{(l)}_{r,m}\|} \left(\left(q_{t,m}\!\!-q^{(l)}_{t,m}\right)\!-\!\left(q_{r,m}\!\!-q^{(l)}_{r,m}\right)\right)\right\} \triangleq
		\widetilde{p}_m,\\
		&\label{A2}\| q_{t,m}\!\!-q_{r,m}\|\!-\!\|q_{t,m} \!\!+ q_{r,m}\|\! \leq\! \|q_{t,m}\!\!-q_{r,m}\| \!-\! \|(q^{(l)}_{t,m}\!\!+q^{(l)}_{r,m}\|\!-\!\textup{Re}\left\{\frac{\left(q^{(l)}_{t,m}\!\!+q^{(l)}_{r,m}\right)^*}{\| q^{(l)}_{t,m}\!\!+q^{(l)}_{r,m}\|} \left(\left(q_{t,m}\!\!-q^{(l)}_{t,m}\right)\!+\!\left(q_{r,m}\!\!-q^{(l)}_{r,m}\right)\right)\right\} \triangleq
		\overline{p}_m.
	\end{align}
	\hrulefill \vspace*{-17pt}
\end{figure*}
where $l$ denotes the number of iterations.
	
Next, considering the possibility that the left-hand side (LHS) of \eqref{P4_C1} is non-concave, we employ the first-order Taylor expansion to obtain its lower bound as \eqref{A3}, shown at the top of the next page, where $\mathbf{q}^{(l)}_{s_i}$ is a given point. Along the same lines, we may replace the LHS of \eqref{P4_C2} with its linear lower bound as \eqref{A4} to ensure the convexity of constraint \eqref{P4_C2}. 
\begin{figure*}[!t]
	\setlength{\belowdisplayskip}{-1pt}
	\normalsize
	\begin{gather}
		\label{A3}\mathbf{q}^H_{s_i} \mathbf{A}_{i,i} \mathbf{q}_{s_i} - \gamma_{th} \sum^{K_I}_{k\ne i}\mathbf{q}^H_{s_i} \mathbf{A}_{i,k} \mathbf{q}_{s_i} 
		\geq 2 \textup{Re}\{\mathbf{q}^H_{s_i} \mathbf{A}_{i,i} \mathbf{q}^{(l)}_{s_i} \}-\left(\mathbf{q}^{(l)}_{s_i}\right)^H \mathbf{A}_{i,i} \mathbf{q}^{(l)}_{s_i}-\gamma_{th}\sum^{K_I}_{k\ne i} \mathbf{q}^H_{s_i}\mathbf{A}_{i,k} \mathbf{q}_{s_i} 
		\triangleq R^{(l)}(\mathbf{q}_{s_i}),\\
		\label{A4}\sum_{k=1}^{K_I} \mathbf{q}^H_{s_j} \mathbf{B}_{j,k} \mathbf{q}_{s_j}\!+\!\sum_{k=1}^{K_E} \mathbf{q}^H_{s_j} \mathbf{C}_{j,k} \mathbf{q}_{s_j} \!\geq \!2 \textup{Re}\left\{\mathbf{q}^H_{s_j} \left(\sum_{k=1}^{K_I}\mathbf{B}_{j,k}+\sum_{k=1}^{K_E}\mathbf{C}_{j,k}\right) \mathbf{q}^{(l)}_{s_j}\right\}\!-\!\left(\mathbf{q}^{(l)}_{s_j}\right)^H\! \!\left(\sum_{k=1}^{K_I}\mathbf{B}_{j,k}\!+\!\sum_{k=1}^{K_E}\mathbf{C}_{j,k}\right)\! \mathbf{q}^{(l)}_{s_j}\!\triangleq\! P^{(l)}\!\!\left(\mathbf{q}_{s_j}\right)\!.
	\end{gather}
	\hrulefill \vspace*{-17pt}
\end{figure*}
As for \eqref{P1_C3}, we may exhibit it further as $\beta^s_{m}=q^*_{s,m}q_{s,m}, s\in \{t,r\}, \forall m \in \mathcal{M}$. Then, following the convex concave procedure (CCP), we can relax it by $\beta^s_{m}\leq q^*_{s,m}q_{s,m} \leq \beta^s_{m}$, the first inequality can be further approximated as $\beta^s_{m}\leq 2 \textup{Re} \{ q^*_{s,m}q^{(l)}_{s,m}\}- q^{(l)*}_{s,m}q^{(l)}_{s,m}$ with its first Taylor expansion. Thus, by introducing new slack variables $\{x_{s,m}\geq 0\}$, problem \eqref{P4} is approximated as 
\begin{subequations}\label{P5}
	\begin{align}
		&\label{P5_C0}	\min_{\mathbf{q}_s,Z_m,X_{s,m}} \ \xi \sum_{m\in\mathcal{M}} Z_m+ \chi \sum_{s\in\{t,r\}}\sum_{m=1}^{2M} X_{s,m}\\ 
		&\label{P5_C1} \!\!{\rm s.t.} \ Z_m \geq \widetilde{p}_m, Z_m \geq \overline{p}_m, \forall m \in \mathcal{M}, \\
		&\label{P5_C2}  \quad R^{(l)} (\mathbf{q}_{s_i}) \geq \gamma_{th} \sigma^2, \forall i \in \mathcal{K_I}, \\ 
		&\label{P5_C3}  \quad P^{(l)} (\mathbf{q}_{s_j}) \geq P_{th}, \forall j \in \mathcal{K_E},\\ 
		&\label{P5_C4}  \quad q^*_{s,m}q_{s,m} \leq \beta^s_{m} + X_{s,m}, \forall m\in \mathcal{M},\\
		&\label{P5_C5} \quad q^{(l)*}_{s,m}q^{(l)}_{s,m}\!\!\!-\!2\textup{Re}\{ q^*_{s,m}q^{(l)}_{s,m}\}\!\leq \!X_{s,m\!+\!M}\!\!-\!\beta^s_m, \forall m \!\in\! \mathcal{M},\\
		&\label{P5_C6}   \quad X_{s,m} \geq 0, m=1,2,\cdots,2M,\\
		&\label{P5_C7}   \quad \eqref{P1_C3},
	\end{align}
\end{subequations}
where $\sum_{s}\sum_{m=1}^{2M}X_{s,m}$ denotes a new penalty function that limits the CCP and $\chi$ represents its associated penalty coefficient. Moreover, in order to improve the convergence performance, we introduce a residual variable vector, denoted by $\Delta\!=\![\delta_1,\cdots,\delta_{K_I}, \delta_{K_I+1},\cdots,\delta_{K_I+K_E}]$. Then, problem \eqref{P5} can be recast to 
\begin{subequations}\label{P6}
	\begin{align}
		&\label{P6_C0}	\min_{\mathbf{q}_s,Z_m,X_{s,m}} -\!\!\sum_{i=1}^{K_I+K_E} \!\!\delta_i \!+\! \xi\!\! \sum_{m\in\mathcal{M}}\!\!Z_m\!\!+\!\! \chi \sum_{s\in\{t,r\}}\sum_{m=1}^{2M} X_{s,m}\\ 
		&\label{P6_C1} {\rm s.t.} \ R^{(l)} (\mathbf{q}_{s_i}) \geq \gamma_{th} \sigma^2+\delta_i,  \forall i \in \mathcal{K_I},\\ 
		&\label{P6_C2} \ \ \quad P^{(l)} (\mathbf{q}_{s_j}) \geq P_{th}+\delta_{j+K_I},  \forall j \in \mathcal{K_E},\\ 
		&\label{P6_C3} \ \  \quad \eqref{P1_C3}, \eqref{P5_C1}, \eqref{P5_C4}, \eqref{P5_C5}, \eqref{P5_C6}.
	\end{align}
\end{subequations}
Note that problem \eqref{P6} is a standard convex problem and can be solved using the CVX solver. However, in order to avoid the conflict between the two penalty terms, the weights between $\chi$ and $\xi$ should be controlled in a reasonable way. In fact, considering that the penlaty term with respect to amplitude constraint typically exerts a greater impact on performance compared to coupled phase-shift constraint, we maintain $\frac{\xi}{\chi}=0.1$ consistently throughout this paper to achieve a higher quality solution.
	
In summary, for given penalty coefficients $\xi$ and $\chi$, by alternately optimizing problems \eqref{P3} and \eqref{P6} several times, we can obtain the suboptimal solution to the original problem \eqref{P2}. Subsequently, by gradually increasing the penalty coefficients and repeating the previous steps until $\sum_{m\in\mathcal{M}} Z_m \leq \epsilon_1$ and $ \sum_{s\in\{t,r\}}\sum_{m=1}^{2M} X_{s,m} \leq \epsilon_2$ are held, the suboptimal solution will converge to a high-quality stationary point. Then, the proposed algorithm can be outlined in \textbf{Algorithm 1}.
\begin{algorithm}[!t]\label{method1}
	\caption{Proposed weight penalty based AO algorithm to solve problem \eqref{P1}.}
	\label{alg:A}
	\begin{algorithmic}[1]
		\STATE {Initialize variables $\{\mathbf{q}^{(0)}_s\}$, penalty coefficients $\xi$ and $\chi$.}
		\REPEAT  
		\STATE {Set the iteration number $l=1$.} 
		\REPEAT 
		\STATE {Solve problem \eqref{P3} with given $\{\mathbf{q}^{(l-1)}_s\}$, update $\{\mathbf{W}^{(l)}_i\}$ and $\{\mathbf{V}^{(l)}_j\}$}.
		\STATE {Solve problem \eqref{P6} with given $\{\mathbf{W}^{(l)}_i\}$ and $\{\mathbf{V}^{(l)}_j\}$ and $\{\mathbf{q}^{(l-1)}_s\}$, update $\{\mathbf{q}^{(l)}_s\}$}.\\
		\STATE Update $l \leftarrow l+1$.\\
		\UNTIL the fractional increase of the objective value is below a threshold $\varepsilon_0$.\\
		\STATE {Update $\mathbf{q}^{(0)}_s$} with the current solutions $\mathbf{q}^{(l)}_s$, and update $\xi$, $\chi$.
		\UNTIL the penalty term is satisfied with predefined convergence condition.
		\STATE {Output} $\mathbf{W}^*_i=\mathbf{W}^{(l)}_i$, $\mathbf{V}^*_j=\mathbf{V}^{(l)}_j$ and $\mathbf{q}^*_s=\mathbf{q}^{(l)}_s$.
	\end{algorithmic}
\end{algorithm}
\section{Proposed Solution for Active STAR-RIS}
Without delving into the intricacies of phase-shift coupling between transmission and reflection, we can directly proceed with beamforming design from the perspective of matrix dimensions. In this case, following the definition in the previous section, we may reformulate the problem \eqref{P1} for the active STAR-RIS as follows:
\begin{subequations}\label{P7}
	\begin{align}
		&\label{P7_C0}\  \min_{{\mathbf{W}_i},{\mathbf{V}_j},\mathbf{Q}_s,\mathbf{A}} \sum_{i\in\mathcal{K_I}}\mathrm{Tr}\left(\mathbf{W}_i\right)+\sum_{j\in\mathcal{K_E}}\mathrm{Tr}\left(\mathbf{V}_j\right)\\ 
		&\label{P7_C1} {\rm s.t.} \mathrm{Tr}\!\left(\mathbf{W}_i\mathbf{H}^H_i\!\mathbf{A}\mathbf{Q}_{s_i}\mathbf{A}\mathbf{H}_i\right)\! \!- \!\gamma_{th}\!\! \!\!\!\!\! \sum_{k\in\mathcal{K_I},k\ne i}\!\!\!\!\!\!\mathrm{Tr}\!\!\left(\mathbf{W}_k\mathbf{H}^H_i\mathbf{A}\mathbf{Q}_{s_i}\mathbf{A}\!\mathbf{H}_i\right)\nonumber \\
		&\ \quad -\gamma_{th}\sigma^2_z\mathrm{Tr}\left(\mathbf{h}^H_{s,i}\mathbf{A}\mathbf{Q}_{s_i}\mathbf{A}\mathbf{h}_{s,i}\right)\!\geq \!\gamma_{th} \sigma^2, \forall i \in \mathcal{K_I}, \\
		&\label{P7_C2} \quad \sum_{i\in \mathcal{K_I}}\!\!\mathrm{Tr}\left(\!\mathbf{W}_i\mathbf{G}_j^H\mathbf{A}\mathbf{Q}_{s_j}\mathbf{A}\mathbf{G}_j\!\right)\!+\!\!\!\sum_{k\in\mathcal{K_E}}\!\!\mathrm{Tr}\left(\!\mathbf{V}_k\mathbf{G}_j^H\mathbf{A}\mathbf{Q}_{s_j}\mathbf{A}\mathbf{G}_j\!\right)\nonumber \\
		&\ \quad +\sigma^2_z\mathrm{Tr}\left(\mathbf{g}^H_{s,j}\mathbf{A}\mathbf{Q}_{s_i}\mathbf{A}\mathbf{g}_{s,j}\right) \! \geq \! P_{th}, \forall j \in \mathcal{K_E},\\
		&\label{P7_C3} \ \quad \mathbf{W}_i \succeq 0, \mathrm{Rank}\left(\mathbf{W}_i\right)=1, \forall i \in \mathcal{K_I},\\
		&\label{P7_C4} \ \quad \mathbf{V}_j \succeq 0, \mathrm{Rank}\left(\mathbf{V}_j\right)=1, \forall j \in \mathcal{K_E}, \\
		&\label{P7_C5} \ \quad \mathbf{Q}_s \succeq 0, \mathrm{Rank}\left(\mathbf{Q}_s\right)=1, \forall s\in\{t,r\},\\
		&\label{P7_C6} \ \quad [\mathbf{Q}_s]_{m,m} \in [0,1], \forall s \in \{t,r\}, \forall m \in \mathcal{M},\\ 
		&\label{P7_C7} \ \quad [\mathbf{Q}_t]_{m,m}+[\mathbf{Q}_r]_{m,m}=1, \forall m \in \mathcal{M},\\ 
		&\label{P7_C8} \ \quad [\mathbf{A}]_{m,m} \in [0,a_{\max}], \forall m \in \mathcal{M},\\ 
		&\label{P7_C9} \ \quad \sum_{s\in\{t,r\}}\!\!\Big(\sum_{i\in \mathcal{K_I}}\!\!\mathrm{Tr}\left(\!\mathbf{W}_i\mathbf{F}^H\mathbf{A}\mathbf{Q}_{s}\mathbf{A}\mathbf{F}\!\right)\!+\!\!\!\nonumber \\
		&  \ \quad\sum_{j\in\mathcal{K_E}}\!\!\mathrm{Tr}\left(\!\mathbf{V}_j
		\mathbf{F}^H\mathbf{A}\mathbf{Q}_{s}\mathbf{A}\mathbf{F}\!\right)\! \!+\! \sigma^2_z\mathrm{Tr}\left(\mathbf{A}\mathbf{Q}_s\mathbf{A}\right) \Big)\!\! \leq\!\! P_a.
	\end{align}
\end{subequations}
It can be observed that the SDR method used to solve problem \eqref{P3} cannot be directly utilized for the solution of this problem \eqref{P7} due to the high coupling of variables. Therefore, we first divide all optimization variables into two blocks, i.e., $\{\mathbf{W}_i$,$\mathbf{V}_j$,$\mathbf{Q}_s\}$ and $\{\mathbf{A}\}$. Based on this foundation, the original problem is accordingly decomposed into two subproblems, which are regarded as the joint beamforming design problem and the amplification feasibility-check problem.
\vspace{-0.25cm}
\subsection{Joint Beamforming Design}
On the one hand, with fixed $\{\mathbf{A}\}$, the joint beamforming design problem is reduced for $\{\mathbf{W}_i$,$\mathbf{V}_j$,$\mathbf{Q}_s\}$. However, the non-convexity of constraints still impedes the solution to the problem. As such, we can rewrite constraints \eqref{P7_C1} and \eqref{P7_C2} in a simpler form as \eqref{A5} and \eqref{A6} according to\cite[\textbf{Lemma 1}]{Mu_star}. In the following, we employ SCA to obtain a suboptimal solution in an iterative manner. For a given point $\{\mathbf{W}^{(l)}_i,\mathbf{V}^{(l)}_j,\mathbf{Q}^{(l)}_s\}$ in the $l$-th iteration, we can obtain the convex upper bound of  \eqref{A5} and \eqref{A6} according to the first Taylor expansion as \eqref{A7} and \eqref{A8}, respectively. Similarly, the non-convexity of the LHS of constraint \eqref{P7_C9} can be approximated by \eqref{A9} in each iteration. As for non-convex rank-one constraints, we still employ the SDR method to ignore them directly. With these transformations, the relaxed problem can be approximated as 
\begin{subequations}\label{P8}
	\begin{align}
		&\label{P8_C0}\  \min_{{\mathbf{W}_i},{\mathbf{V}_j},\mathbf{Q}_s} \sum_{i\in\mathcal{K_I}}\mathrm{Tr}\left(\mathbf{W}_i\right)+\sum_{j\in\mathcal{K_E}}\mathrm{Tr}\left(\mathbf{V}_j\right)\\
		&\label{P8_C1} \quad \quad {\rm s.t.}\quad \ [R_{\text{act},i}]^{\textup{ub}} \leq 0, \forall i \in \mathcal{K_I}, \\
		&\label{P8_C2} \quad \quad \quad \quad \ \ [P_{\text{act},j}]^{\textup{ub}} \leq 0, \forall j \in \mathcal{K_E}, \\
		&\label{P8_C3} \quad \quad \quad \quad \ \ \mathbf{W}_i \succeq 0, \forall i \in \mathcal{K_I}, \\
		&\label{P8_C4} \quad \quad \quad \quad \ \ \mathbf{V}_j \succeq 0, \forall j \in \mathcal{K_E}, \\
		&\label{P8_C5} \quad \quad \quad \quad \ \ \mathbf{Q}_s \succeq 0, \forall s \in \{t,r\}, \\
		&\label{P8_C6} \quad \quad \quad \quad \ \  [P_{\textup{c}}]^{\textup{ub}} \leq P_S, \\
		&\label{P8_C7} \quad \quad \quad \quad \ \ \eqref{P7_C6}, \eqref{P7_C7}, \eqref{P7_C8}.
	\end{align}
\end{subequations}
This is a standard SDP problem and can be solved by CVX. Note that the rank-one constraint for AP beamforming is always satisfied, while for the STAR-RIS, we can reconstruct it using the Gaussian randomization procedure \cite{SDR}, the details of which are omitted here for brevity.
\begin{figure*}[!t]
	\setlength{\belowdisplayskip}{-1pt}
	\normalsize
	\begin{gather}
		R_{\textup{act},i}\triangleq\frac{1}{2}\|\mathbf{W}_i-\mathbf{H}^H_i\!\mathbf{A}\mathbf{Q}_{s_i}\mathbf{A}\mathbf{H}_i\|^2_{F}-\frac{1}{2}||\mathbf{W}_i||^2_F-\frac{1}{2}\|\mathbf{H}^H_i\!\mathbf{A}\mathbf{Q}_{s_i}\mathbf{A}\mathbf{H}_i\|^2_F \nonumber \\
		\label{A5} +\gamma_{th}\!\!\!\!\sum_{k\in\mathcal{K_I},k\ne i}\!\!\!\left(\frac{1}{2}\|\mathbf{W}_k\!+\!\mathbf{H}^H_i\!\mathbf{A}\mathbf{Q}_{s_i}\mathbf{A}\mathbf{H}_i\|^2_{F}\!-\!\frac{1}{2}||\mathbf{W}_k||^2_F\!-\!\frac{1}{2}\|\mathbf{H}^H_i\!\mathbf{A}\mathbf{Q}_{s_i}\mathbf{A}\mathbf{H}_i\|^2_F\right)\!+\!\gamma_{th}\sigma^2_z\mathrm{Tr}\left(\mathbf{h}^H_{s,i}\mathbf{A}\mathbf{Q}_{s_i}\mathbf{A}\mathbf{h}_{s,i}\right)\!+\!\gamma_{th} \sigma^2\!\leq \! 0,\\
		P_{\textup{act},j}\triangleq\sum_{i\in \mathcal{K_I}}\left(\frac{1}{2}\|\mathbf{W}_i-\mathbf{G}^H_j\!\mathbf{A}\mathbf{Q}_{s_j}\mathbf{A}\mathbf{G}_j\|^2_{F}-\frac{1}{2}||\mathbf{W}_i||^2_F-\frac{1}{2}\|\mathbf{G}^H_i\!\mathbf{A}\mathbf{Q}_{s_j}\mathbf{A}\mathbf{G}_j\|^2_F \right)\nonumber \\
		\label{A6}+\sum_{k\in\mathcal{K_E}}\left(\frac{1}{2}\|\mathbf{V}_k-\mathbf{G}^H_j\mathbf{A}\mathbf{Q}_{s_j}\mathbf{A}\mathbf{G}_j\|^2_{F}-\frac{1}{2}||\mathbf{V}_k||^2_F-\frac{1}{2}\|\mathbf{G}^H_j\!\mathbf{A}\mathbf{Q}_{s_j}\mathbf{A}\mathbf{G}_j\|^2_F\right)-\sigma^2_z\mathrm{Tr}\left(\mathbf{g}^H_{s,j}\mathbf{A}\mathbf{Q}_{s_j}\mathbf{A}\mathbf{g}_{s,j}\right)+P_{th}\!\leq 0.
	\end{gather}
	\hrulefill \vspace*{-17pt}
\end{figure*}
\begin{figure*}[!t]
	\setlength{\belowdisplayskip}{-1pt}
	\normalsize
	\begin{gather}
		R_{\textup{act},i}\!\leq\!\frac{1}{2}\|\mathbf{W}_i\!-\!\mathbf{H}^H_i\!\mathbf{A}\mathbf{Q}_{s_i}\mathbf{A}\mathbf{H}_i\|^2_{F}\!+\!\frac{1}{2}||\mathbf{W}^{(l)}_i||^2_F\!-\!\mathrm{Tr}\left(\left(\mathbf{W}^{(l)}_i\right)^H\!\!\!\mathbf{W}_i\right)\!+\!\frac{1}{2}\|\mathbf{H}^H_i\!\mathbf{A}\mathbf{Q}^{(l)}_{s_i}\mathbf{A}\mathbf{H}_i\|^2_F\nonumber \\ -\mathrm{Tr}\left(\left(\mathbf{A}\mathbf{H}_i\mathbf{H}^H_i\!\mathbf{A}\mathbf{Q}^{(l)}_{s_i}\mathbf{A}\mathbf{H}_i\mathbf{H}^H_i\mathbf{A}\right)^H
		\mathbf{Q}_{s_i}\right)+\gamma_{th}\!\!\!\!\sum_{k\in\mathcal{K_I},k\ne i}\!\!\!\bigg(\frac{1}{2}\|\mathbf{W}_k\!+\!\mathbf{H}^H_i\!\mathbf{A}\mathbf{Q}_{s_i}\mathbf{A}\mathbf{H}_i\|^2_{F}\!+\!\frac{1}{2}||\mathbf{W}^{(l)}_k||^2_F\!-\mathrm{Tr}\left(\left(\mathbf{W}^{(l)}_i\right)^H\mathbf{W}_i\right) \nonumber \\
		\label{A7}+\!\frac{1}{2}\|\mathbf{H}^H_i\!\mathbf{A}\mathbf{Q}^{(l)}_{s_i}\mathbf{A}\mathbf{H}_i\|^2_F\!-\!\mathrm{Tr}\left(\left(\mathbf{A}\mathbf{H}_i\mathbf{H}^H_i\!\mathbf{A}\mathbf{Q}^{(l)}_{s_i}\mathbf{A}\mathbf{H}_i\mathbf{H}^H_i\mathbf{A}\right)^H\mathbf{Q}_{s_i}\right)\bigg)
		\!+\!\gamma_{th}\sigma^2_z\mathrm{Tr}\left(\mathbf{h}^H_{s,i}\mathbf{A}\mathbf{Q}_{s_i}\mathbf{A}\mathbf{h}_{s,i}\right)\!+\!\gamma_{th} \sigma^2\! \triangleq\![R_{\text{act},i}]^{\textup{ub}},\\
		P_{\textup{act},j}\!\leq\sum_{i\in \mathcal{K_I}}\bigg(\frac{1}{2}\|\mathbf{W}_i-\mathbf{G}^H_j\!\mathbf{A}\mathbf{Q}_{s_j}\mathbf{A}\mathbf{G}_j\|^2_{F}+\frac{1}{2}||\mathbf{W}^{(l)}_i||^2_F-\!\mathrm{Tr}\left(\left(\mathbf{W}^{(l)}_i\right)^H\!\!\!\mathbf{W}_i\right)+\frac{1}{2}\|\mathbf{G}^H_j\!\mathbf{A}\mathbf{Q}^{(l)}_{s_j}\mathbf{A}\mathbf{G}_j\|^2_F \nonumber \\
		-\mathrm{Tr}\left(\left(\mathbf{A}\mathbf{G}_j\mathbf{G}^H_j\!\mathbf{A}\mathbf{Q}^{(l)}_{s_j}\mathbf{A}\mathbf{G}_j\mathbf{G}^H_j\mathbf{A}\right)^H
		\mathbf{Q}_{s_j}\right) \bigg) +\sum_{k\in\mathcal{K_E}}\bigg(\frac{1}{2}\|\mathbf{V}_k-\mathbf{G}^H_j\mathbf{A}\mathbf{Q}_{s_j}\mathbf{A}\mathbf{G}_j\|^2_{F}+\frac{1}{2}||\mathbf{V}^{(l)}_k||^2_F-\!\mathrm{Tr}\left(\left(\mathbf{V}^{(l)}_k\right)^H\!\!\!\mathbf{V}_k\right) \nonumber \\
		\label{A8}+\frac{1}{2}\|\mathbf{G}^H_j\!\mathbf{A}\mathbf{Q}^{(l)}_{s_j}\mathbf{A}\mathbf{G}_j\|^2	-\mathrm{Tr}\left(\left(\mathbf{A}\mathbf{G}_j\mathbf{G}^H_j\!\mathbf{A}\mathbf{Q}^{(l)}_{s_j}\mathbf{A}\mathbf{G}_j\mathbf{G}^H_j\mathbf{A}\right)^H
		\mathbf{Q}_{s_j}\right) \bigg)-\sigma^2_z\mathrm{Tr}\left(\mathbf{g}^H_{s,j}\mathbf{A}\mathbf{Q}_{s_j}\mathbf{A}\mathbf{g}_{s,j}\right)+P_{th} \triangleq [P_{\text{act},j}]^{\textup{ub}}.
	\end{gather}
	\hrulefill \vspace*{-17pt}
\end{figure*}
\begin{figure*}[!t]
	\setlength{\belowdisplayskip}{-1pt}
	\normalsize
	\begin{gather}
		[P_{\textup{c}}]^{\textup{ub}} \triangleq \sum_{s\in\{t,r\}}\ \bigg( \sum_{i\in \mathcal{K_I}} 	\bigg(\frac{1}{2}\|\mathbf{W}_i+\mathbf{F}^H\!\mathbf{A}\mathbf{Q}_{s}\mathbf{A}\mathbf{G}\|^2_{F}+\frac{1}{2}||\mathbf{W}^{(l)}_i||^2_F-\!\mathrm{Tr}\left(\left(\mathbf{W}^{(l)}_i\right)^H\!\!\!\mathbf{W}_i\right)+\frac{1}{2}\|\mathbf{F}^H\!\mathbf{A}\mathbf{Q}^{(l)}_{s}\mathbf{A}\mathbf{F}\|^2_F \nonumber \\
		-\mathrm{Tr}\left(\left(\mathbf{A}\mathbf{F}\mathbf{F}^H\!\mathbf{A}\mathbf{Q}^{(l)}_{s}\mathbf{A}\mathbf{F}\mathbf{F}^H\mathbf{A}\right)^H
		\mathbf{Q}_{s}\right) \bigg)+\sum_{k\in\mathcal{K_E}}\bigg(\frac{1}{2}\|\mathbf{V}_k+\mathbf{F}^H\mathbf{A}\mathbf{Q}_{s}\mathbf{A}\mathbf{F}\|^2_{F}+\frac{1}{2}||\mathbf{V}^{(l)}_k||^2_F-\!\mathrm{Tr}\left(\left(\mathbf{V}^{(l)}_k\right)^H\!\!\!\mathbf{V}_k\right) \nonumber \\
		\label{A9}+\frac{1}{2}\|\mathbf{F}^H\!\mathbf{A}\mathbf{Q}^{(l)}_{s}\mathbf{A}\mathbf{F}\|^2_F	-\mathrm{Tr}\left(\left(\mathbf{A}\mathbf{F}\mathbf{F}^H\!\mathbf{A}\mathbf{Q}^{(l)}_{s}\mathbf{A}\mathbf{F}\mathbf{F}^H\mathbf{A}\right)^H
		\mathbf{Q}_{s}\right)+\sigma^2_z\mathrm{Tr}\left(\mathbf{A}\mathbf{Q}_s\mathbf{A}\right) \bigg)
		\bigg).
	\end{gather}
	\hrulefill \vspace*{-17pt}
\end{figure*}
\subsection{Amplification Feasibility-Check}
On the other hand, with fixed $\{\mathbf{W}_i,\mathbf{V}_j,\mathbf{Q}_s\}$, the original problem is reduced to a feasibility-check problem with amplification matrix $\{\mathbf{A}\}$. Specially, since all $\{\mathbf{W}_i,\mathbf{V}_j,\mathbf{Q}_s\}$ are semipositive definites, we can define $\widetilde{\mathbf{W}}_i\triangleq\mathbf{W}^{\frac{1}{2}}_i$, $\widetilde{\mathbf{V}}_j\triangleq\mathbf{V}^{\frac{1}{2}}_j$, $\widetilde{\mathbf{Q}}_s\triangleq\mathbf{Q}^{\frac{1}{2}}_s$. Then, the feasibility-check problem is shown as
\begin{subequations}\label{P9}
	\begin{align}
		&\label{P9_C0}\  \textup{Find} \quad \mathbf{A} \\
		&\label{P9_C1}{\rm s.t.} \ \|\widetilde{\mathbf{W}}_i\mathbf{H}^H_i\mathbf{A}\widetilde{\mathbf{Q}}_{s_i}\|^2_F-\gamma_{th}\!\!\!\sum_{k\in\mathcal{K_I},k\ne i}\|\widetilde{\mathbf{W}}_k\mathbf{H}^H_i\mathbf{A}\widetilde{\mathbf{Q}}_{s_i}\|^2_F \nonumber \\
		&\quad \ -\gamma_{th}\sigma^2_z\|\mathbf{h}^H_{s,i}\mathbf{A}\widetilde{\mathbf{Q}}_{s_i}\|^2_F \geq \gamma_{th} \sigma^2, \forall i \in \mathcal{K_I}, \\
		&\label{P9_C2}\quad \ \sum_{i\in\mathcal{K_I}}\|\widetilde{\mathbf{W}}_i\mathbf{G}^H_j\mathbf{A}\widetilde{\mathbf{Q}}_{s_j}\|^2_F+\!\!\!\sum_{k\in\mathcal{K_E}}\|\widetilde{\mathbf{V}}_k\mathbf{G}^H_j\mathbf{A}\widetilde{\mathbf{Q}}_{s_j}\|^2_F, \nonumber \\
		&\quad \ +\sigma^2_z\|\mathbf{g}^H_{s,j}\mathbf{A}\widetilde{\mathbf{Q}}_{s_i}\|^2_F \geq P_{th}, \forall j \in \mathcal{K_E}, \\
		&\label{P9_C3}\quad \ \sum_{s \in \{t,r\}} \bigg(\sum_{i\in \mathcal{K_I}}\|\widetilde{\mathbf{W}}_i\mathbf{F}^H\mathbf{A}\widetilde{\mathbf{Q}}_{s}\|^2_F+\sum_{j\in\mathcal{K_E}}\|\widetilde{\mathbf{V}}_j\mathbf{F}^H\mathbf{A}\widetilde{\mathbf{Q}}_{s}\|^2_F \nonumber \\
		&\quad \ + \sigma^2_z \|\mathbf{A}\widetilde{\mathbf{Q}}_s\|^2_F \big) \leq P_a, \\
		&\label{P9_C4}\quad \ \eqref{P7_C8}.
	\end{align}
\end{subequations}
For the LHS of non-convex constraints \eqref{P9_C1} and \eqref{P9_C2}, we can obtain their lower bound using their first Taylor expansion expression, as shown in \eqref{A10} and \eqref{A11} at the top of the next page. Consequently, problem \eqref{P9} can be approximated as a standard convex SDP as follows:
\begin{subequations}\label{P10}
	\begin{align}
		&\label{P10_C0}\  \textup{Find} \quad \mathbf{A} \\
		&\label{P10_C1}{\rm s.t.} \  [R_{\text{act},i}]^{\textup{lb}} \geq \gamma_{th}, \forall i \in \mathcal{K_I}, \\
		&\label{P10_C2}\quad \ \ [P_{\text{act},j}]^{\textup{lb}} \geq P_{th}, \forall j \in \mathcal{K_E},\\
		&\label{P10_C3}\quad \ \ \eqref{P7_C8}, \eqref{P9_C3}.
	\end{align}
\end{subequations}
To solve this simpler problem, CVX is utilized. To sum up, the intractable problem \eqref{P1} for the active STAR-RIS is first reformulated as a SDP problem \eqref{P7}. Then, the SCA framework is applied to alternately solve the two approximated subproblems \eqref{P8} and \eqref{P10} decomposed from \eqref{P7}. Since the optimal goal only exists in \eqref{P8}, which is non-increasing, and \eqref{P10} is a feasibility-check problem, the convergence of the proposed algorithm is guaranteed. Additionally, the details of the proposed algorithm are summarized in \textbf{Algorithm 2}.
\begin{algorithm}[!t]\label{method2}
	\caption{Proposed iterative algorithm to solve \eqref{P7}.}
	\label{alg:A}
	\begin{algorithmic}[1]
		\STATE {Initialize variables $\{\mathbf{W}^{(0)}_i,\mathbf{V}^{(0)}_j,\mathbf{Q}^{(0)}_s\}$, and $\{\mathbf{A}^{(0)}\}$, set the iteration number $l=1$.}
		\REPEAT 
		\STATE {Solve problem \eqref{P8} with given $\{\mathbf{A}^{(l-1)}\}$, update $\{\mathbf{W}^{(l)}_i,\mathbf{V}^{(l)}_j, \mathbf{Q}^{(l)}_s\}$} according to SCA framework.
		\STATE {Solve problem \eqref{P10} with given $\{\mathbf{W}^{(l)}_i,\mathbf{V}^{(l)}_j, \mathbf{Q}^{(l)}_s\}$, update $\{\mathbf{A}^{(l)}\}$} according to SCA framework.\\
		\STATE Update $l \leftarrow l+1$.\\
		\UNTIL the fractional increase of the objective value is below a threshold $\varepsilon_0$.\\
		\STATE {Output} $\mathbf{W}^*_i\!\!=\!\mathbf{W}^{(l)}_i$, $\mathbf{V}^*_j\!\!=\!\mathbf{V}^{(l)}_j$, $\mathbf{Q}^*_s\!\!=\!\mathbf{Q}^{(l)}_s$, and $\mathbf{A}^*\!\!=\!\mathbf{A}^{(l)}$.
	\end{algorithmic}
\end{algorithm}
\begin{figure*}[t]
	\setlength{\belowdisplayskip}{-1pt}
	\normalsize
	\begin{gather}
		\label{A10}\!-\!\|\widetilde{\mathbf{W}}_i\mathbf{H}^H_i\mathbf{A}^{(l)}\widetilde{\mathbf{Q}}_{s_i}\|^2_F\!+\!2\mathrm{Tr}\left(\left(\mathbf{H}_i\mathbf{W}_i\mathbf{H}_i^H\!\!\mathbf{A}^{(l)}\mathbf{Q}_{s_i}\right)^H\!\!\! \mathbf{A}\right)\!-\!\gamma_{th}\!\!\!\!\!\sum_{k\in\mathcal{K_I},k\ne i}\!\!\!\!\|\widetilde{\mathbf{W}}_k\mathbf{H}^H_i\mathbf{A}\widetilde{\mathbf{Q}}_{s_i}\|^2_F\!-\!\gamma_{th}\sigma^2_z\|\mathbf{h}^H_{s,i}\mathbf{A}\widetilde{\mathbf{Q}}_{s_i}\|^2_F\!\triangleq\! [R_{\text{act},j}]^{\textup{lb}},\\
		-\|\widetilde{\mathbf{W}}_i\mathbf{G}^H_i\mathbf{A}^{(l)}\widetilde{\mathbf{Q}}_{s_j}\|^2_F+2\mathrm{Tr}\left(\left(\mathbf{G}_j\mathbf{W}_i\mathbf{G}_j^H\mathbf{A}^{(l)}\mathbf{Q}_{s_j}\right)^H \mathbf{A}\right)-\|\widetilde{\mathbf{V}}_j\mathbf{G}^H_j\mathbf{A}^{(l)}\widetilde{\mathbf{Q}}_{s_j}\|^2_F+2\mathrm{Tr}\left(\left(\mathbf{G}_j\mathbf{V}_j\mathbf{G}_j^H\mathbf{A}^{(l)}\mathbf{Q}_{s_j}\right)^H \mathbf{A}\right)\nonumber \\
		\label{A11}-\sigma^2_z\left(\|\mathbf{g}^H_{s,j}\mathbf{A}^{(l)}\widetilde{\mathbf{Q}}_{s_j}\|^2_F-2\mathrm{Tr}\left(\left(\mathbf{g}_{s,j}\mathbf{g}_{s,j}^H\mathbf{A}^{(l)}\mathbf{Q}_{s_j}\right)^H \mathbf{A}\right)\right)\triangleq [P_{\text{act},j}]^{\textup{lb}}.
	\end{gather}
	\hrulefill \vspace*{-17pt}
\end{figure*}
\subsection{Computational Complexity Analysis}
On the one hand, the original problem \eqref{P1} for the passive STAR-RIS is solved by a two-layer algorithm exploiting penalty method. In the outer layer, the penalty coefficients are updated in each iteration. While in the inner layer, the reformulated problem \eqref{P2} with fixed $\xi$ and $\chi$ is decomposed into a standard SDP and a vector dimension optimization problem. Of them, the former can be solved by the interior point method \cite{SDR}, whose computational complexity is determined by $\mathcal{O}\left(KN^{3.5}\right)$, and the computational complexity of the latter is determined by $\mathcal{O}\left(4M^2+6M\right)$. Thus, the approximate computational complexity of \textbf{Algorithm 1}  is given by $\mathcal{O}^{\textup{pass}}=\mathcal{O}\left(I_{\textup{out}}I_{\textup{in}}\left(KN^{3.5}+4M^{2}+6M\right)\right)$, where $I_{\textup{out}}$ and $I_{\textup{in}}$ denote the number of outer and inner loops, respectively, and $K=K_I+K_E$. On the other hand, for the active STAR-RIS, the original problem is divided into two subproblems with a standard SDP form, which can also be optimized by the interior point method iteratively. Therefore, let $I_{\textup{act}}$ denote the number of loops, the approximate computational complexity of \textbf{Algorithm 2} is given by $\mathcal{O}^{\textup{act}}=\mathcal{O}\left(I_{\textup{act}}\left(KN^{3.5}+3M^{3.5}\right)\right)$.
\section{Numerical Results}
To comprehensively compare the performance of passive and active STAR-RISs in SWIPT systems, we consider two different resource constraints. First, we compare them under the same size, where the power consumption of the hardware is fixed by $M$. In this scenario, the power consumption of the active STAR-RIS may be much higher than that of the passive STAR-RIS. Thus, to ensure a fair comparison, we add the power consumption $P_S$ of the STAR-RIS to the power consumed $P_T$ by the AP as in \cite{Gao_active}. Second, we investigate the comparison with the same STAR-RIS power budget $P_S$. In this case, the optimal number of elements for the passive STAR-RIS can be directly determined as $M^{*}=\frac{P_S}{p_c}$, while the $M^{*}$ for the active STAR-RIS needs to be further explored. Based on the above analysis, the numerical results in this paper are developed from different perspectives.
\subsection{Simulation Setup}
\begin{figure}[t]
	\setlength{\abovecaptionskip}{0cm}   
	\setlength{\belowcaptionskip}{-0.18cm}   
	\setlength{\textfloatsep}{7pt}
	\centering
	\includegraphics[width=2.4in]{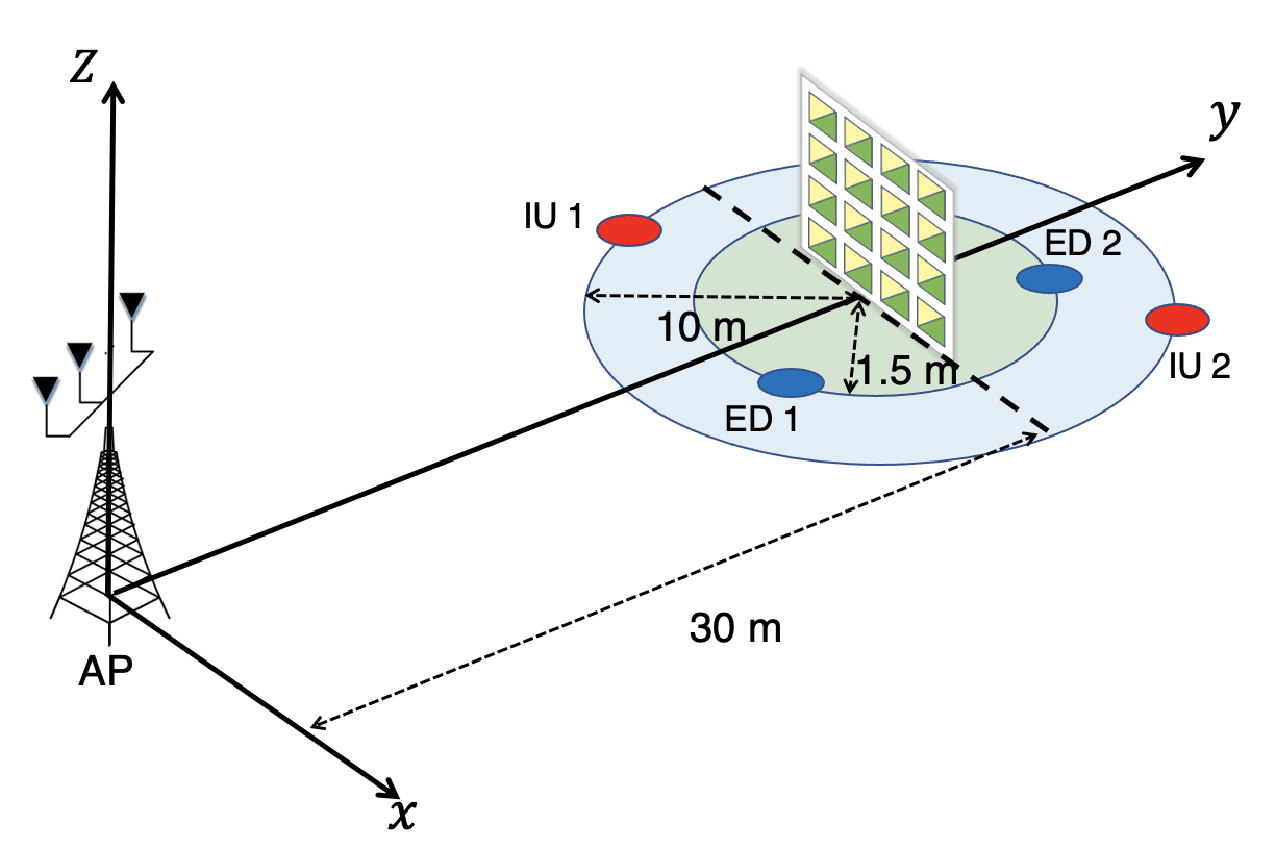}
	\caption{Simulation setup}
	\label{simulation}
\end{figure}
We consider a simulation setup in the 3D coordinate as shown in Fig. \ref{simulation}, where the AP is located at $[0,0,6 \textup{m}]^T$, while the STAR-RIS is deployed at a location 30 meters away from the AP and its central coordinates are $[0,30\textup{m},2\textup{m}]^T$. The EDs and IUs are randomly located on circles centered at the STAR-RIS with a radius of $r_E=1.5$ m and $r_I = 10$ m. In particular, we consider a simple yet widely applicable user distribution, consisting of one ED and one IU in each of the T and R regions. Besides, we assume that the AP is equipped with $N=4$ antennas and the communication operates in the $f_c=20$ GHz band, then $\lambda_c=0.015$ m, $d=\frac{\lambda_c}{2}=0.0075$ m. For the other parameters, unless otherwise noted, they are set as follows: $P_{\textup{c}}=0.1$ mW, $P_{\textup{b}}=0.3167$ mW, $a_{\max}^2=40$ dB \cite{Long_active}, $L=10$, $\eta=0.8$, $\sigma^2=\sigma^2_z=-100$ dBm, $\varepsilon_0=10^{-5}$, and $\epsilon_1=\epsilon_2=10^{-3}$.
	
In addition, three baseline schemes are considered: 1) \textbf{Passive STAR-RISs (Independent)}: In this scheme, the phase-shift of $\textbf{q}_t$ and $\textbf{q}_r$ in each element can be adjusted from $[0,2\pi)$ independently. 2) \textbf{Conventional Active/Passive RISs}: In this scheme, two $M/2$-element conventional reflecting-only RIS and transmitting-only RIS are deployed adjacent to each other. 3) \textbf{Passive STAR-RISs (Element-wise):} In this scheme, the coupled phase-shift constraint in passive STAR-RISs is solved by the element-wise method \cite{Liu_passive_model}.
\subsection{Convergence of  Proposed Algorithms}
\begin{figure}[t]
	\setlength{\abovecaptionskip}{0cm}   
	\setlength{\belowcaptionskip}{0cm}   
	\setlength{\textfloatsep}{7pt}
	\centering
	\includegraphics[width=2.4in]{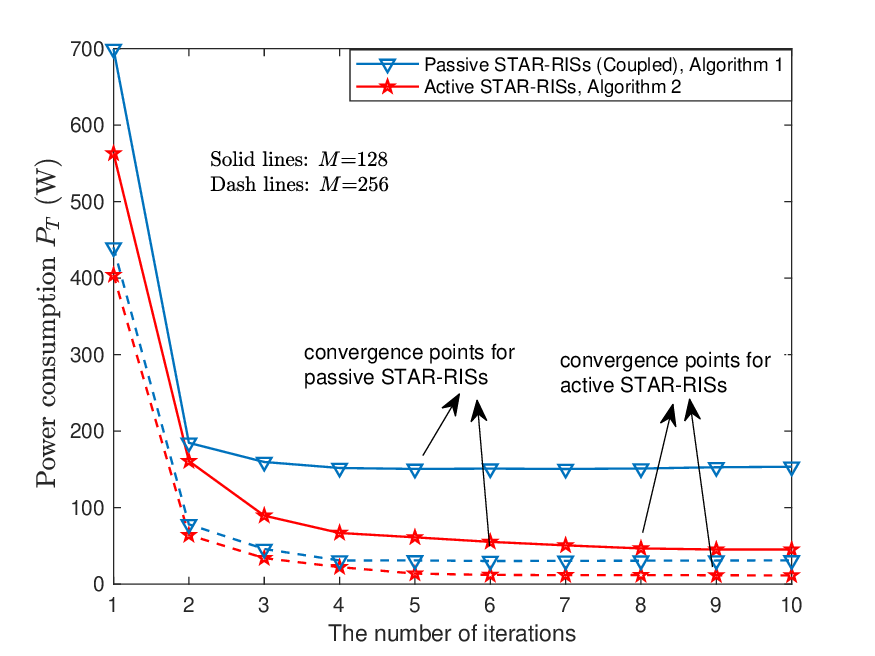}
	\caption{Convergence behavior of propsoed Algorithms}
	\label{convergence}
\end{figure}
In Fig. \ref{convergence}, we show the convergence behavior of the proposed algorithms for both passive and active STAR-RISs. Note that the number of iterations here for \textbf{Algorithm 1} refers to the number of outer loops. The results depict that all curves fly down as the number of iterations increases and eventually stabilize at a fixed value. However, this process takes longer for active STAR-RISs compared to passive STAR-RISs. This is because active STAR-RISs require additional optimization of the amplification factor matrix, which is not present in passive STAR-RISs. Besides, \textbf{Algorithm 1} optimizes phase-shift from a vector dimension, whereas \textbf{Algorithm 2} uses a matrix dimension. Nevertheless, in any case, the convergence rates remain within 10 iterations even if $M=256$. This confirms the feasibility of our proposed algorithms.

\subsection{Performance Comparison Under the Same Size}
In this subsection, the performance comparison is based on the same number of STAR-RIS elements. To ensure fairness, we also consider the power consumption of the STAR-RIS as a reference for the comparison. In this case, the power consumption $P_S$ of the passive STAR-RIS is determined by $P_S=MP_c$, while the power consumption $P_S$ of the active STAR-RIS is determined by $P_S=M(P_c+P_b)+P_a$.

\begin{figure}[t]
	\setlength{\abovecaptionskip}{-0.05cm}   
	\setlength{\belowcaptionskip}{-0.1cm}   
	\setlength{\textfloatsep}{7pt}
	\centering
	\vspace{-0.18cm}
	\includegraphics[width=2.4in]{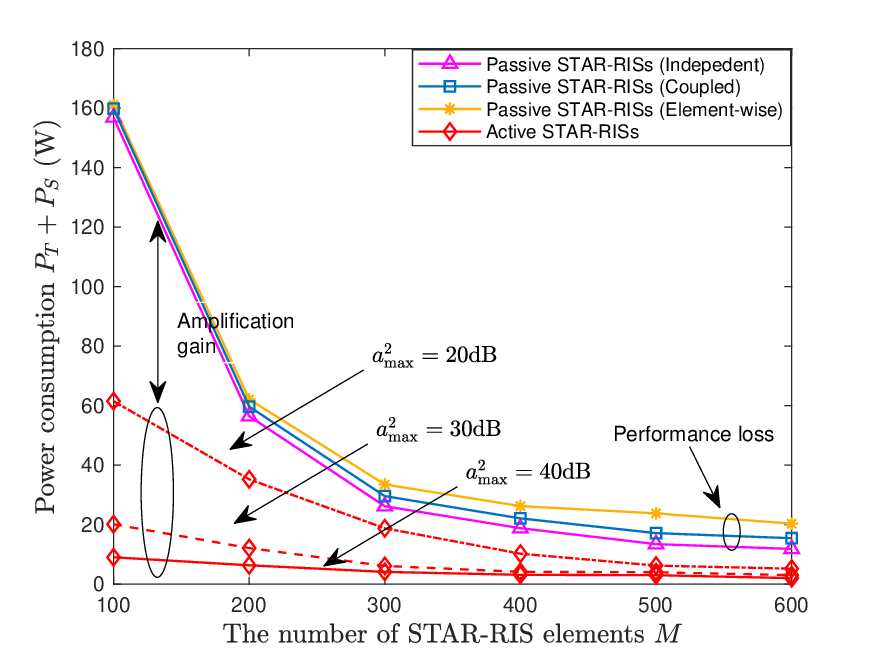}
	\caption{$P_T+P_S$ versus the number of STAR-RIS elements}
	\label{res3}
\end{figure}
\emph{1) Power Consumption Versus STAR-RIS Elements}: In Fig. \ref{res3}, we explore the power consumption of $P_T+P_S$ versus the number of STAR-RIS elements. We set the $P_a=0.1$ W, QoS requirements of IUs and EDs as $R_{th}=2$ bits/s/Hz and $P_{th}=2$ $\mu$W. It can be observed that active STAR-RISs have a significant advantage over passive STAR-RISs when the $M$ is small, which is due to the amplification gain provided by active load. However, as $M$ increases, the performance gap is decreasing. This can be explained as follows. Under the same QoS, the DoF gain from increasing $M$ helps passive STAR-RISs effectively mitigate the double fading effect, thus reducing the transmit power significantly. However, in active STAR-RISs, the increase of $M$ not only compresses the amplification capability of each element but also introduces additional hardware overheads, which inhibit the performance enhancement of the system by the additional DoFs. The same reasons can be used to explain the slowing growth rate of active STAR-RISs performance as the maximum allowable amplification factor $a_{\max}$ increases. Regarding the performance loss due to coupled phase-shift in passive STAR-RISs, it is almost negligible when $M$ is small. Although this loss increases with a larger $M$, it remains relatively limited, indicating the robustness of our proposed algorithm. In addition, compared to the element-wise method, our proposed algorithm demonstrates significant performance improvement. This advantage is due to the optimization of the entire STAR-RIS as a whole, rather than optimizing each element separately.

\begin{figure}[t]
	\setlength{\abovecaptionskip}{-0cm}   
	\setlength{\belowcaptionskip}{-0.1cm}   
	\setlength{\textfloatsep}{7pt}
	\centering
	\subfigbottomskip=-0.7pt
	\subfigcapskip=-7pt 
	\subfigure[For different harvested power requirements $P_{th}$.]{\label{versus_P}
		\includegraphics[width= 2.4in]{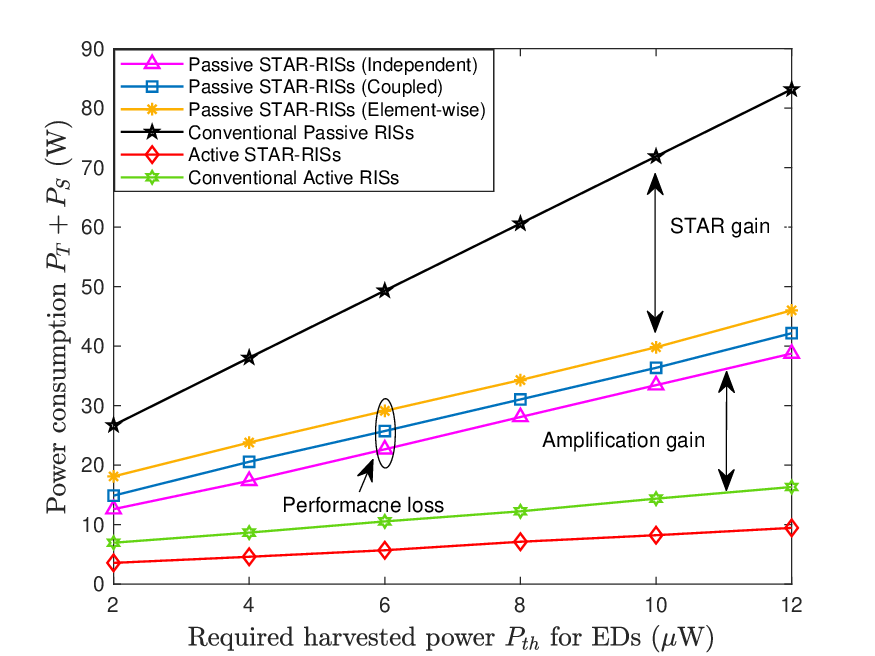}}
	\subfigure[For different data rate requirements $R_{th}$.]{\label{versus_R}
		\includegraphics[width= 2.4in]{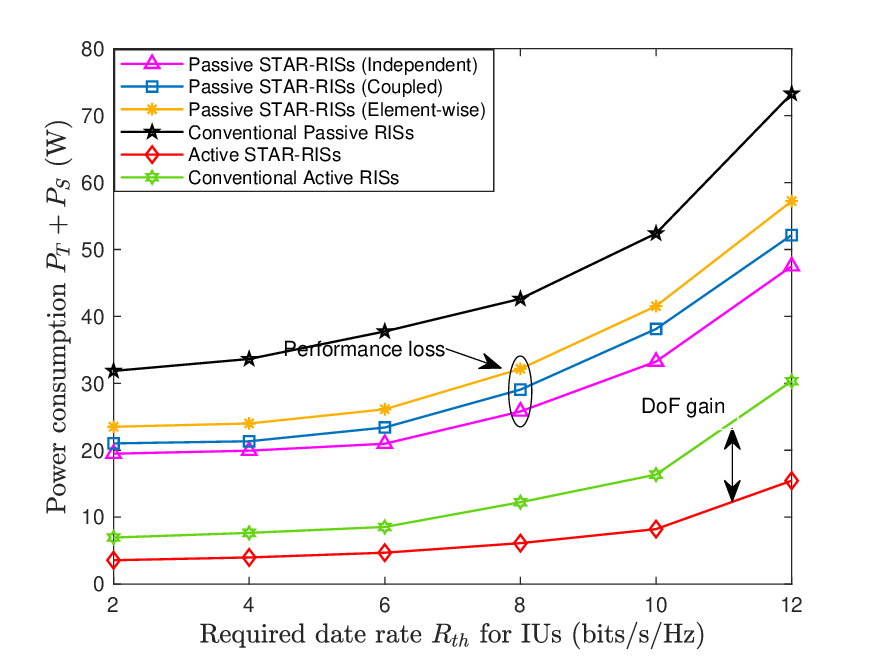}}
	\caption{\textcolor{black}{$P_T+P_S$ versus QoS requirements.} } 
	\label{versus_QoS}
\end{figure}
\emph{2) Power Consumption Versus QoS Requirements}: Fig. \ref{versus_QoS} plots the power consumption of $P_T+P_S$ versus the QoS requirements with $M=30\times20$ and $P_a=0.1$ W. In this case, EDs are located in the near-field region while IUs are located in the far-field region of the STAR-RIS. Specially, we investigate the power consumption versus $P_{th}$ with fixed $R_{th}=2$ bits/s/Hz in Fig. \ref{versus_P}. As can be observed, as the $P_{th}$ increases, the performance gap between STAR-RISs and conventional RISs is widening, especially in the passive model comparison. This is largely due to the additional DoFs from the STAR-RIS. However, this phenomenon is less significant in the active model because the amplification gain has a more substantial impact on the system than the influence of external DoFs. This also explains why conventional active RISs outperform passive STAR-RISs in this scenario. In addition, to cope with the increase in $P_{th}$, active STAR-RISs are able to control the cost of $P_S$ more effectively, thanks to the signal amplification gain and thermal noise gain from the active load. 

A similar situation is observed in Fig. \ref{versus_R}, which plots the power consumption versus the $R_{th}$ with fixed $P_{th}=4$ $\mu$W. The difference is that the performance loss caused by coupled phase-shift in this figure is more pronounced, especially at high data rates. This can be attributed to the following reasons: First, IUs in the far-field region receive plane waves from the AP, necessitating greater beamforming directionality. Second, a higher $R_{th}$ increases challenge of beamforming in terms of enhancing useful signal and suppressing interference, particularly when users are located on both sides of STAR-RISs.
\subsection{Performance Comparison with the Same Power Budget}
With the same power budget $P_S$, the dimensions of passive STAR-RISs and active STAR-RISs can differ significantly. In particular, the optimal solution for passive STAR-RISs is achieved by using up all the budget to extend its size, i.e., $M^*=\frac{P_S}{P_c}$. However, the optimal size of active STAR-RIS is to be determined. Therefore, we first explore the performance comparison versus the number of active STAR-RIS elements, and then investigate the effect of different channel models due to STAR-RIS size differences in the compraison.

\begin{figure}[t]
	\setlength{\abovecaptionskip}{-0cm}   
	\setlength{\belowcaptionskip}{-0.1cm}   
	\setlength{\textfloatsep}{7pt}
	\centering
	\subfigbottomskip=-0.5pt
	\subfigcapskip=-7pt 
	\subfigure[For different STAR-RIS power budget $P_S$.]{\label{versus_active1}
		\includegraphics[width= 2.4in]{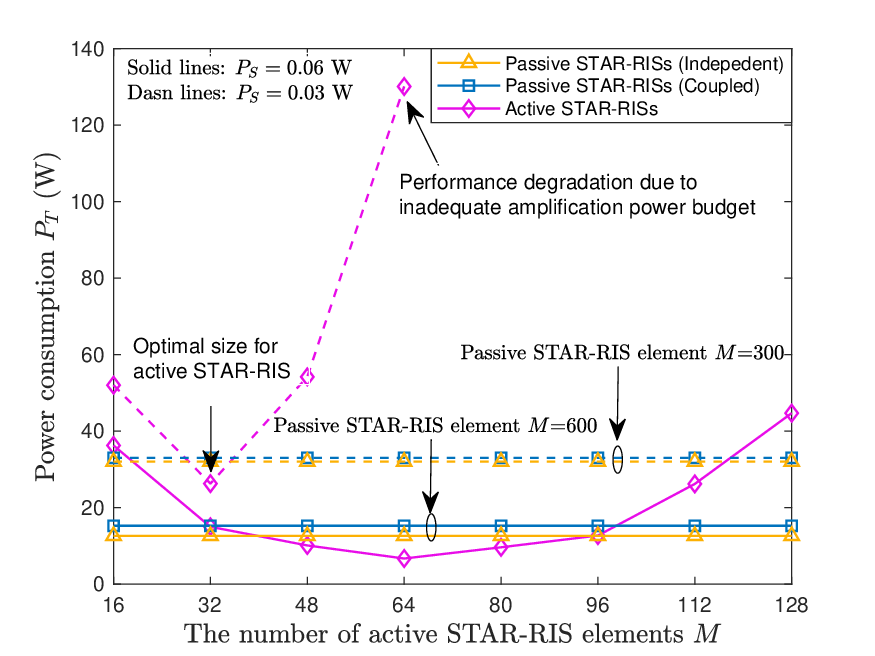}}
	\subfigure[For different user QoS requirement $R_{th}$ and $P_{th}$.]{\label{versus_active2}
		\includegraphics[width= 2.4in]{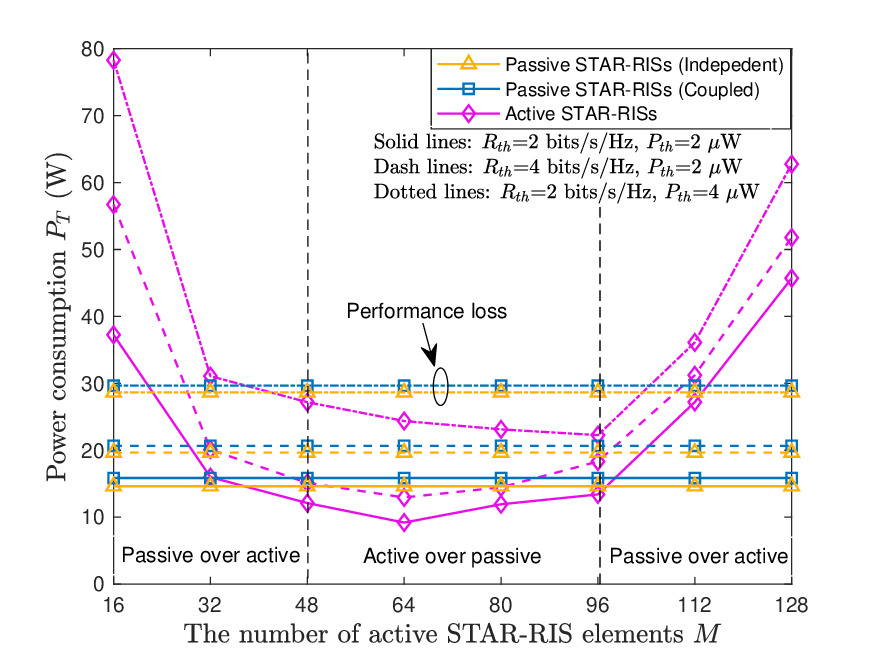}}
	\caption{\textcolor{black}{$P_T$ versus the number of active STAR-RIS elements.} } 
	\label{versus_active}
\end{figure}
\emph{1) Power Consumption Versus Number of Active STAR-RIS Elements:}
In Fig. \ref{versus_active}, we study the AP power consumption $P_T$ versus the number of active STAR-RIS elements. Particularly, the results presented in Fig. \ref{versus_active1} are based on different power budgets $P_S$ when QoS requirements are fixed with $R_{th}=2$ bits/s/Hz and $P_{th}=2$ $\mu$W. It can be observed that the AP power consumption in passive STAR-RISs assisted systems is constant, whereas that in active STAR-RISs assisted systems initially decreases and then increases as $M$ grows. Furthermore, the latter outperforms the former only when the active STAR-RISs are configured at a moderate size. This can be explained as follows. Given the power budget, the optimal $M^*$ for passive STAR-RISs is fixed with $\frac{P_S}{P_c}$. While for active STAR-RISs, the power budget must accommodate both hardware design and amplifier control. Consequently, when $M$ is small, there may be a surplus of power due to the limitation of $a_{\max}$. By contrast, when $M$ is large, the excessive hardware consumption significantly reduces the power available for amplification, thereby degrading the system performance. As a result, this trade-off poses a significant challenge for hardware design implementation in practice.

In Fig. \ref{versus_active2}, we extend the above study for different QoS requirements with $P_S=0.03$ W, which present similar results to Fig. \ref{versus_active1}. In addition, the optimal number of elements for active STAR-RISs increases as $P_{th}$ increases. This is because the thermal noise and DoF gains achieved from adding a certain amount of elements outweigh the loss due to the reduced amplification power budget. However, this phenomenon is not significant among noise-sensitive IUs.

\begin{figure}[t]
	\setlength{\abovecaptionskip}{-0cm}   
	\setlength{\belowcaptionskip}{-0.1cm}   
	\setlength{\textfloatsep}{7pt}
	\centering
	\subfigbottomskip=-1.9pt
	\subfigcapskip=-7pt 
	\subfigure[Beam pattern for active STAR-RIS.]{\label{versus_far}
		\includegraphics[width= 2.4in]{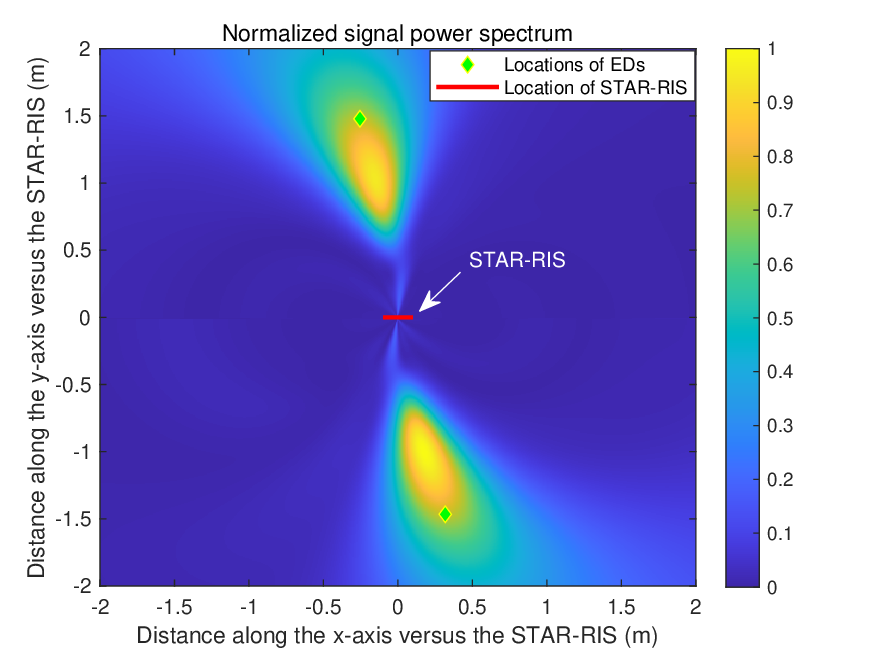}}
	\subfigure[Beam pattern for passive STAR-RIS.]{\label{versus_near}
		\includegraphics[width= 2.4in]{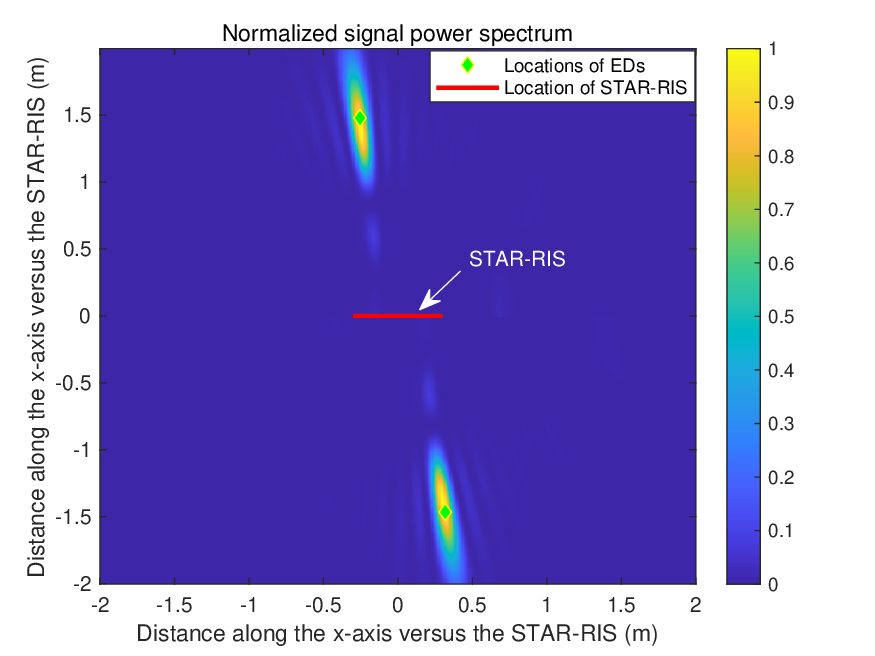}}
	\caption{\textcolor{black}{Beam pattern comparison under same power budget.} } 
	\label{versus_model}
\end{figure}
\emph{2) Beam Pattern Comparison with Same Power Budget:}
In Fig. \ref{versus_model}, we compare the beam pattern for active and passive STAR-RISs under the same power budget, where we consider $P_{S}\!=\!0.06$ W, $R_{th}\!=\!2$ bits/s/Hz and $P_{th}\!=\!2$ $\mu$W. To maximize system efficiency, we assume that the active STAR-RIS is configured with a number of elements of $M=8\times 8$, where all users are located in the far-field region. For passive STAR-RISs, the optimal size with  given power budget is $M=600$, if we make $M_x=30$ and $M_z=20$, then its Rayleigh distance will be determined as $\frac{2D^2}{\lambda_c}=9.75$ m, where EDs will fall into the near-field region while the IUs are distributed in the far-field region. Note that since IUs are located in the far-field region in both scenarios and their harvested energy magnitude is much lower than that of EDs, we mainly focus on the beam pattern comparison of EDs. As such, Fig. \ref{versus_far} plots the beam pattern for EDs in far-field region, where the power intensity at the EDs is at a very high level. However, its signal power exhibits a gradual decrease in the propagation path, which suggests that the  beam steering in far-field causes some energy dissipation. By contrast, Fig. \ref{versus_near} illustrates the beam pattern of EDs in the near-field. It can be seen that beam focusing in this scenario can accurately converge the energy into specific locations of EDs, thus increasing the efficiency of WPT.

\section{Conclusions}
In this paper, a STAR-RIS assisted SWIPT system was studied, where both active and passive STAR-RISs were considered for comparison. In this context, the minimum power consumption optimization problem was formulated for both models. To solve the resulting intractable problem, a weight penalty based AO algorithm was proposed for passive STAR-RISs, and an iterative method exploiting SCA and convex optimization techniques was utilized in active STAR-RISs. Numerical results unveiled that STAR-RISs can significantly reduce power consumption for the AP by utilizing a small power budget, confirming their potential in future SWIPT applications. Additionally, given the same aperture size, active STAR-RISs outperform passive ones, especially at the smaller aperture size. However, under the same power budget, passive STAR-RISs become a more robust option due to their simpler design, which avoids the complex trade-offs between hardware power consumption and amplification power budget required by active STAR-RISs. Thus, addressing these design challenges for active STAR-RISs will be the focus of our future work to fully leverage their potential.
\bibliographystyle{IEEEtran} 
\bibliography{a_v_p_bib}
\end{document}